\documentclass[12 pt,
preprint,
nofootinbib,
 amsmath,amssymb,
]{revtex4-2}

\usepackage{graphicx,amssymb,amsfonts,amsmath}
\usepackage{dcolumn}
\usepackage{bm}
\usepackage[mathlines]{lineno}

\newcommand{\ket}[1]{|#1\rangle}
\newcommand{\bra}[1]{\langle #1|}
\newcommand{\braket}[2]{\langle #1|#2\rangle}

\usepackage{comment}

\usepackage{xcolor}

\newcommand{\Tr}{\operatorname{Tr}}

\begin{document}
\title{Successes and challenges of using Semidefinite Programming for the study of Spin Chain Hamiltonians}

\author{David Berenstein}
 \email{dberens@physics.ucsb.edu}
 \affiliation{Department of Physics, UC Santa Barbara 93106}
 
\author{P. N. Thomas Lloyd}%
 \email{plloyd@ucsb.edu}
\affiliation{Department of Physics, UC Santa Barbara 93106}%

\date{\today}

\begin{abstract}
We study semidefinite programming (SDP) methods to analyze spin chain Hamiltonians. 
We examine the ground state energy, the first excited charged states and ground state correlators in two simple models: the Ising model in a transverse magnetic field and the closely related 3-state Potts model. Our goal is to understand precisely what the SDP program is doing and when it works well, why it does so. We focus on the following novel ingredients: using charge constraints to obtain excited states and to see if additional constraints from integrable models are effective at improving the method.
 At criticality we also explore to what extent we can use approximate Virasoro correlators to extract conformal data: the central charge and some critical exponents of charged states. We also use these to identify the location of the phase transition.  In the special case where the system is made of free fermions we prove that the SDP finds the exact energy of the ground state and produces the correct two point functions of the fermions.
Away from free fermion theories, the SDP gets progressively worse at estimating data beyond the value of the ground state energy (like correlation functions), although it qualitatively matches these. 
In order to be effective, the SDP  seems to run into scaling issues where the amount of input needed scales poorly with the lattice volume.
\end{abstract}

\maketitle



\section{Introduction}

Recent work on many body quantum mechanics (including lattice quantum field theories) has explored the possibility that semidefinite programming (SDP) methods can be used to find the ground state energy and its correlators. In some cases the excited states of such systems are also accessible \cite{barthel2012solving,Han:2020bkb,Lawrence:2021msm}. Some of these methods are related to the numerical bootstrap program for conformal field theory (this program is reviewed in \cite{Poland:2018epd}) and also carry that moniker.
A useful collection of results in semidefinite methods for a variety of physics problems can be found in \cite{tavakoli2024semidefinite}.

The main idea of this program is to exploit universal facts of general quantum mechanical systems to make inroads. The first axiom to exploit (one can think of it as a feature of the problem)  is that the inner product on the Hilbert space is positive definite. This is also known as unitarity. In that sense, expressions like the following can be shown to be positive
\begin{equation}
    \bra \psi   {\cal O}^\dagger  {\cal O} \ket \psi = || {\cal O} \ket \psi||^2\geq 0 \label{eq:positivity1}
\end{equation}
If $\ket \psi$  also happens to be an eigenstate of the Hamiltonian with energy $E$, then 
\begin{equation}
\hat H \ket \psi = E\psi\label{eq:energy_eig}
\end{equation}
and this can be shown to produce additional relations between expectation values, such as 
\begin{equation}
\bra \psi [\hat H, {\cal O}] \ket \psi=0 \label{eq:dynamics}
\end{equation}
The idea is that since the numerical value of  $E$ might be hard to calculate exactly from first principles, we might want to minimize the expectation value of $E$ subject to a subset of the positivity constraints like \eqref{eq:positivity1}, and also subject to the dynamical relations (constraints) derived from equations \eqref{eq:energy_eig}, \eqref{eq:dynamics}. Such a program, because we are minimizing $E$, would be able to find the ground state energy, or more precisely a lower bound on the energy of the ground state, but not necessarily the excited states of the system. On the other hand, the variational method produces upper bounds on the energy of the ground state. A combination of both can in principle determine $E$ to high accuracy, so long as both bounds are close to the true value. In that sense, at the very worst, these methods based on positivity are complementary to other approaches.

Although the description above might seem somewhat abstract as a program, there are simple analytical models where this method will lead to solutions of the exact complete spectrum of a system, or more generally the list of unitary irreducible representations of a Lie algebra. This is the familiar use of ladder operators for the study of representation theory of operator algebras in quantum mechanics.  For example, the harmonic oscillator and the matrices of angular momentum are solved using positivity in the Hilbert space plus the commutation relations of the ladder operators with the Hamiltonian.
In the angular momentum case it is commutations with the $z$ component of angular momentum $L_z$ that do the trick. This is well known physics and is standard in most textbooks in quantum mechanics. The key is that in most quantum systems one has a stability property: that $E$ is actually bounded from below. In the case of the harmonic oscillator, since the Hamiltonian is a sum of squares, the energy is bounded below by zero. 
One can then use descent relations to show that a lowering operator will not only decrease $E$, but that it will eventually lead to a failure of unitarity unless a termination condition is found (these are relations like  $a\ket 0=0$). The spectrum can then be generated from $\ket 0$ (or more generally a highest weight state) by acting multiple times with ladder operators on that state.

Another related approach is to fix $E$, that is, to assume that its value is known exactly and numerically equal to $E$. One then checks consistency:  that the positivity constraints are satisfied, given all the other dynamical constraints. In that setup, one searches for ``islands of positivity" where one can guarantee that the value of $E$ selected is not ruled out by the finite subset of constraints that have been implemented numerically. Increasing the list of constraints will narrow the possibilities, that is, the size of the islands decreases as one includes more information \footnote{This is the approach that is most similar in spirit to the numerical conformal bootstrap, which has been incredibly successful in the task of  finding critical exponents of non-trivial conformal field theories \cite{El-Showk:2012cjh}.}.
This second route has been shown to be successful empirically in certain quantum mechanical systems for one particle in one dimension (see the early works \cite{Han:2020bkb,Berenstein:2021dyf,Bhattacharya:2021btd}). In these cases there are enough constraints and the problem is recursive enough, so that solving the problem at fixed $E$ requires knowing only a finite amount of initial data. From that data, all expectation values of polynomials in position, momenta $x,p$ can be derived. 
The most successful method to calculate the values of this set of data results from optimizing the positive quantities so that they are as positive as possible \cite{Berenstein:2022unr} (the precise algorithm maximizes the minimal eigenvalue of a putative positive matrix, which is also an SDP problem). After that is done, one can search in $E$ to find in what regions of the values of $E$, that particular auxiliary minimal eigenvalue is actually positive. Using this algorithm it is possible to obtain exact mathematical bounds 
for the possible values of the  eigenvalues of the Hamiltonian. In these setups the list of positive quantities is in principle infinite, but the truncation to fit it in a computer gives a finite SDP problem.

If we go to the problem of minimizing $E$, or some situations where $E$ is known, the constraints can also be used to study the ground state properties of some general quantum mechanics problems, like those that appear in matrix quantum mechanics \cite{Lin:2023owt}. 

It is possible to generalize these ideas to statistical mechanics of lattice systems and matrix integrals \cite{Anderson:2016rcw,Kazakov:2021lel}.
In special situations,  it is possible to prove exponentially fast convergence to the correct solution on a small subset of these problems, which can be used to solve matrix models via the numerical construction of their associated orthogonal polynomials \cite{Berenstein:2025itw,Berenstein:2026wky}. These proofs of convergence all depend on the precise details of the problems that are being studied and do not seem to generalize easily to other setups.

When studying Hamiltonian lattice systems like a spin chain, we can ask if this positivity bootstrap method is an effective   route to obtain the correct physics of the ground state, and maybe even the excited states as well. Here, when we say effective, we mean it in the sense of algorithmic complexity. We would like to answer whether the algorithm converges fast enough to the correct values of the physical quantities and with a computational cost that scales well with the size of the problem. The results in this direction are mixed and do not seem to enjoy the numerical advantages that arise in the other setups described above. Unfortunately, we cannot answer this most relevant question. Instead, we show in simple examples that generically the problem of finding the ground state of a spin chain by semidefinite programming  methods is a difficult problem (in the sense of computational resources needed). The one exception being when the system is equivalent to free fermions and the basis of free fermion operators is chosen.

Some of our previous work has shown that sometimes it is possible to extract critical properties of some lattice models \cite{georgeberensteinmeoldpaper}, by applying bootstrap methods to finite-size spin chains and extracting conformal data like the central charge and some critical exponents through finite-size scaling analysis.  Also, \cite{Nancarrow:2022wdr} 
developed bootstrap bounds on energy gaps in quantum spin systems. These works establish the feasibility of using relaxed optimization constraints to study critical phenomena, though they primarily focus on ground state energies rather than the broader spectrum of excited states.

Previous work has noted that optimization procedures naturally discover (or more precisely, prefer) fermionic representations \cite{georgeberensteinmeoldpaper}, suggesting that SDP methods can identify the underlying mathematical structures that simplify some quantum many-body problems. However, systematic studies of how additional physical constraints affect this emergence of preferred degrees of freedom and what information can be extracted beyond ground state properties remain limited.

The goal of this paper is to push these spin chain systems further and to see if additional information that is known about these spin chain systems can be implemented numerically within the SDP framework. 
In this paper we extend the SDP framework for quantum spin chain analysis by implementing three classes of additional constraints in the analysis. First, we introduce charge constraints based on the global symmetries of spin chain models, enabling access to some excited states in different charge sectors. These charge constraints enable the calculation of energy gaps and the identification of excited states beyond the ground state.
Second, we incorporate constraints derived from conserved quantities (which we derive from the existence of a boost operator) when the system is integrable.  We are using these ideas to investigate whether additional conserved quantities obtained from these additional conserved  operators improve the accuracy of physical observables. Also, since many of these systems have non-invertible symmetries at criticality, we can study how additional constraints affect  how accurately the SDP method is in  representing these particular symmetries (here we mean that Ward identities for the symmetry are satisfied). 
Third, we study approximate Virasoro algebra constraints using Fourier-transformed Hamiltonians following the Koo-Saleur prescription \cite{KOO1994459}. These  provide a way to identify possible continuous phase transitions associated with a conformal field theory. We also explore the utility of such Virasoro constraints to find a method to directly extract additional conformal data from finite-size lattice systems. 
We apply and investigate these enhanced methods to the transverse-field Ising model and three-state Potts model, comparing our results to exact analytical predictions (or exact diagonalization on small systems) and exploring the limitations/successes of the SDP approach. 

  We analyze how the SDP optimization procedure  discovers anticommutation relations and fermionic variable structures when presented with appropriate constraints and a suitable basis of operators. We analyze the mathematical origins of this emergence and its relationship to the optimization landscape. In a sense, the SDP does not know the full fermionic algebra, so there is a question of how many of the known properties of such systems is present numerically in the SDP data that can be extracted from a solution of the SDP problem.
  
  We conclude that the SDP seems to be challenged in its use, meaning that it seems to be exponentially hard to get good data beyond the objective function (the ground state energy). In particular, extracting good data on correlators of the ground state is very difficult, except on situations where the system is a free field theory. Even in integrable models that are not described by free fermions the challenges remain after one adds the integrability constraints. Qualitatively, on the other hand, the method seems to give a reasonable picture of the correlators when enough data is encoded in the SDP.

It seems clear that to make further progress, it will be necessary to add additional physical information to improve the method. The ability to incorporate physical principles as optimization constraints suggests that, at least in principle, new avenues for studying quantum field theories using these SDP methods are possible and should be further studied. We similarly caution and present arguments as to why an SDP approach may be fundamentally limited in its scope and necessitates further input such as those seen in matrix product states (entanglement and renormalization group), or the conformal bootstrap, where there is additional symmetry that can be exploited to obtain information and further constrain the system. 

The paper is organized as follows. In section \ref{sec:prel} we discuss some basic aspects of how semidefinite programming problems arise in the study of quantum systems. We also show convergence of the semidefinite programming method to the minimum of the Hamiltonian, provided that a complete basis of operators is used. We also pose the problem of what happens when one tries to take the thermodynamic limit and the issues arising when one truncates (relaxes) the semidefinite programming problem that we actually want to solve. In section \ref{sec:models} we introduce the two models we will solve by semidefinite programming (bootstrap) methods.
These are the Ising model in a transverse magnetic field and an analogous clock model with ${\mathbb Z}_3$ symmetry that at criticality is the 3-state Potts model.
The first one is integrable for all coupling constant values, while the second one is integrable only at criticality. We study the ground state and first excited charged state for both problems and in the second case we also compare with exact diagonalization methods for small lattices.  We  implement integrable charge constraints to see if they improve the SDP result. We also implement the Koo-Saleur prescription for Virasoro algebra generators to see if some expectation values related to these  allow for a better identification of the critical point in the SDP.
In section \ref{sec:free} we further explore the Ising model and we show why on any free fermion system the SDP will converge to the correct ground state if only the two fermion correlation functions are used to optimize.
We show that as soon as the problem becomes non-linear, the advantage of the free fermion basis gets lost and the spin chain problem becomes hard to solve with SDP methods. A similar parfermion basis for the 3 state Potts model do not solve the system, although it helps to show that the non-invertible symmetry Ward identities at criticality are satisfied by the solution. Finally in section \ref{sec:conc}, we conclude with our observations and future directions.

\section{Preliminaries: Understanding the quantum mechanical Semidefinite Programming (bootstrap) algorithm.}\label{sec:prel}

The  problem that we want to solve in general is finding the minimal eigenvalue of a Hamiltonian: given a Hamiltonian $H$ (for the purposes of this article we will work in a finite dimensional space), we want to find the energy of the ground state. That is, the minimal value of $\langle \hat H \rangle$ over all possible states. In other problems, one also would like to determine the full spectrum of $H$. In this paper  $H$ will be  the Hamiltonian of a spin chain. In that case,  various arguments suggest that finding the spectrum becomes too hard, but the ground state is accessible to an SDP approach. 

Here, a state will be implicitly defined by a density matrix $\rho$, so that $\Tr(\rho)=1$ and $\rho\succeq 0$. We are finding the solution to  $\min(\Tr(\rho H))|_{\rho\succeq 0}$ that satisfy positivity and $\Tr(\rho)=1$. If we constrain $\rho$ to a subset of the possible $\rho$, then this would be a particular case of  a variational problem approach to solve the system. Our algorithm for finding bounds will be distinct, as follows from the quantum mechanical bootstrap program. To understand that algorithm, we need some additional information.

The set of $\rho\succeq 0$ is the set of (semidefinite) positive Hermitian matrices. It is a convex cone set. The cut $\Tr(\rho)=1$ is a linear constraint on a convex set. That makes the set of density matrices into a convex set as well. It is, however,  not a cone any longer. 

The positivity of this set is also characterized by the following property: given a positive operator ${\cal P}\succeq 0$, then it must be the case that $\Tr(\rho {\cal P})\geq 0$.

The functional $\Tr(\rho {\cal O})=\langle {\cal O}\rangle$ is a linear function on the set of operators ${\cal O}$ and therefore belongs to the dual space of the operators, which is also a linear space. Using the non-degenerate operator metric $||{\cal O}||^2= \Tr({\cal O}^\dagger {\cal O} )$, we can think of $\rho$ as an operator itself because the norm lets us identify the vector space of operators with the dual vector space (here we are using the finite dimension of the operator algebra to do this identification). Such a matrix can be diagonalized and results in $\rho$ being positive as follows. We diagonalize $\rho = \sum \rho_\alpha \ket \alpha \bra \alpha$. It is trivial to show that ${\cal O}_\alpha=\ket \alpha \bra \alpha \succeq 0$, and then using $\Tr(\rho {\cal O}_\alpha)\geq 0 $ we find that $\rho_\alpha\geq 0$. Hence $\rho$ is a positive linear combination of positive matrices, which is also positive. The normalization condition $\Tr(\rho)=1= \sum \rho_\alpha$ guarantees that the $\rho_\alpha$ have a probabilistic interpretation, which is common in how the density matrix is interpreted in quantum mechanics. 

The quantum mechanical bootstrap assumes the following. There exists a state in the Hilbert space $\psi$ such that $\bra \psi H \ket \psi= E$ realizes the minimum eigenvalue of $H$. In that state, the following constraints apply
\begin{eqnarray}
 \hat H \ket \psi & =& E \ket \psi\\
 \bra \psi [ \hat H, {\cal O}] \ket \psi &=&0
\end{eqnarray}
where ${\cal O}$ is an arbitrary operator.
Thought this way, the state is an eigenstate of the given energy, and there is a list of expectation values that vanish. 
Turning this into the properties of $\rho$, we state that 
\begin{eqnarray}
    \Tr(\rho \hat H) &=& E\\
    \Tr( \rho {\cal P}) &\geq &0 \label{eq:positivity}\\
    \Tr( \rho [H, {\cal O}]) &=& 0 \label{eq:cons}
\end{eqnarray}
The third line of  constraints are linear constraints on $\rho$, so they perform linear cuts in the set of $\rho$ and these preserve the convexity of the space of $\rho$ that solve the constraints. The positivity constraints are convex: convex linear combinations of $\rho$ preserves them. We could also implement $\Tr(\rho {\cal O} \hat H) = E \Tr(\rho {\cal O})$, but in spin chain models $H$ is not a local operator, so this becomes numerically expensive. This is why equation \eqref{eq:cons} is preferred for these systems.

The last constraint is equivalent (if we take it over all operators ${\cal O}$) to $[\rho,H]=0$. This means that $\rho$ is diagonal in the same basis as $H$ (both can be diagonalized simultaneously). This can be used to show that in the full problem we must have that $E=\sum \rho_\alpha E_\alpha$ in that same mutual basis. Then, we find the refined constraints $\rho_\alpha\geq 0$ and $\sum \rho_\alpha=1$. It is then easy to show that $E\geq \min(E_\alpha)$, and when we minimize $E$, that $E=\min(E_\alpha)$. Barring degeneracy, $\rho=\ket E \bra E$.
Otherwise, $\rho$ is an arbitrary density matrix of the degenerate subspace of minimal energy.

When we don't solve the problem in the full algebra, the quantity $\Tr(\rho \hat H)= E$ is treated as the objective function to minimize, given a subset of the constraints. Since the function $\Tr(\rho \hat H)$ is linear in $\rho$, it is a particular case of a convex function.
The positivity constraint \eqref{eq:positivity} is preserved under convex combinations, so even in the relaxed setting, the abstract space of  functionals $\rho$ compatible with the subset of the constraints that we choose is still convex. It is also important to remember that we only impose positivity with \eqref{eq:positivity}, but not directly with $\rho$. Positivity of $\rho$ is an end product if we work with the full algebra of operators, but not with a subset. Hence, a partial solution of a truncated problem does not imply that a $\rho\succeq 0$ exists. If one can find such a $\rho$, let us call it $\rho^*$, then the solution $\rho^*$ must be (one of) the exact answer(s) to the full problem. 
One proves it the following way: such a $\rho$ would satisfy all positivity constraints (even the ones not implemented) so one can think of $\rho^*$ as a solution of the same problem within some variational approach. That way, it would furnish an upper bound on the minimum energy. However, the same bound would be a lower bound from the relaxed problem (all possible $\rho$ would have higher energy because the relaxed problem produces lower bounds on $E$, as we will show later). Combining the two gives a sandwich argument proof: we thus get the exact energy of the ground state.  

Since the set of $\rho$ is convex itself, we are trying to minimize a convex function over a convex set. Notice also that in the case of a finite dimensional operator algebra, the space of $\rho$ positive with $\Tr(\rho)=1$ is actually also a compact set ($\rho$ is bounded in norm by $1$), so the problem is guaranteed to have a solution. The relaxed problem might run-off to $-\infty$, but it is expected that this will be cured by having enough of the positive operators ${\cal P}$ as semidefinite constraints. What this means is that whatever negativity is implicit in a non-fully positive $\rho$ is mild.

So far, we have talked about the set of $\rho \succeq 0$ as if we have access to the full set of constraints given by \eqref{eq:positivity} and \eqref{eq:cons}. Since we are considering a spin chain with $N$ sites, and the operator algebra grows in dimension as $\exp(\alpha N)$, the full problem with all of the constraints is of exponential complexity.
The Hamiltonian Bootstrap algorithm is essentially to impose only some of these constraints and use the partial information to get bounds on 
$E$, hopefully with only polynomial resources on the volume.

Hence the problem of finding $E$ (or bounds on $E$) becomes a problem in convex optimization. In the original formulation in terms of a state
$\ket \psi$ one is minimizing over a compact set, but not a convex set. The state $\ket \psi$ are unit vectors in the Hilbert space and this space is not convex: it is a complex projective space. It is in passing to a density matrix formulation that the problem becomes a convex optimization problem, because with density matrices we can take convex combinations and obtain an allowed density matrix.

Convex optimization problems are very useful: it is not only guaranteed that a solution to that local minimization problem of density matrices in a compact set exists, but that the solution is essentially unique (or in the case of degeneracy, the solution is also a convex set).

The way to  set up the problem is to define a subset of operators ${\cal O}_\gamma$ and take the span of these operators so that a representative operator of this set can be written as  ${\cal O}=\sum c_\gamma {\cal O}_\gamma$. From ${\cal O}$ we can build a positive operator 
\begin{equation}
   {\cal  P} = {\cal O}^\dagger {\cal O}
\end{equation}
The constraint $\Tr(\rho {\cal P}) \geq 0 $ can be rephrased as a constraint on the expectation values $M_{\alpha\beta}=\Tr( \rho  {\cal O}_\alpha^\dagger {\cal O}_\beta)$ given by stating that the matrix $M$ is positive semidefinite
\begin{equation}
M \succeq 0
\end{equation}
Now, if $H\in Span( {\cal O}^\dagger_\alpha {\cal O}_\beta)$, meaning $H=\sum C^{\alpha\beta} {\cal O}_\alpha^\dagger {\cal O}_\beta)$  then the problem of computing $\langle H\rangle $ can be written as a linear combination of the $M_{\alpha\beta} $ as follows
\begin{equation}
    \langle H \rangle = \Tr( C M)\equiv C^{\alpha \beta} M_{\alpha \beta}
\end{equation}
and since $H$ is self-adjoint, then $C$ is also self-adjoint.

The subset ${M\succeq 0} $ is a subset of the 
positive constraints and the set of possible positive semidefinite $M$ matrices is also a convex cone. 
The problem $\min( \Tr( C M))|_{M\succeq 0}$ is again a convex optimization problem that has a solution $E_*$. Also, if some of the linear combinations $[H,\tilde {\cal O}]$ can be expressed in terms of the linear space $Span( {\cal O}^\dagger_\beta{\cal O}_\alpha)$, these become additional linear constraints on the set of possible solutions $M$. We call this method {\em relaxing} the constraints defined by \eqref{eq:positivity}, and we impose as many of the equalities \eqref{eq:cons} as is possible within this set.

The output of the SDP is both the value of $E_*$ and a specific $M$ that satisfies all the given constraints and for which $\Tr( C M))=E_*$.
That is, the program gives a number for each $M_{\alpha\beta}=\langle {\cal O}^\dagger_\alpha{\cal O}_\beta\rangle$, so it also estimates the correlators of the problem (expectation values of operators).

It is easy to show that the minimum value $E_*$ of this auxiliary problem satisfies $E_*\leq E$. The idea is that given the solution $\rho$ to the original problem, one can show that $M_\rho$ and $E$ satisfy all the constraints of this relaxed problem. Hence $M_\rho$ defines a point of the configuration space over which we are minimizing. This way, because $E^*$ is a true minimum, the relaxed  problem provides a lower bound $E_*\leq E$ and an approximation to $M_\rho$.
As we increase the size of the basis ${\cal O}_\alpha$ (we assume that they are linearly independent as they are being generated), every time we add new elements to the basis keeping all the old generators, we find that the larger problem satisfies $E_{*,small} \leq E_{*,bigger}\leq E$, so that we approximate $E$ from below and the problem converges weakly.
It converges because in the full problem we get the exact $E$. We say that it converges weakly because we do not seem in general to be able to do better than the sandwiched inequalities $E_{*,small} \leq E_{*,bigger}\leq E$.

The proof of these pair of inequalities goes as follows: $M_{bigger,*}$ induces values of $M_{smaller}$ by evaluation on the solution, and that value satisfies all the constraints of the smaller problem. Hence, just as $M_\rho$ was used before, we are using the induced matrix $M_{induced}$ to make the argument. Basically, we are using information that $M_{smaller}$ is a true minimum and $M_{induced}$ satisfies the constraints of the smaller problem. This is the argument we were using to argue that if one can prove that a $\rho\succeq 0$ exists for some $E^*$ solution to a relaxed problem, then that is the exact answer.

As soon as the set  $Span( {\cal O}^\dagger_\alpha {\cal O}_\beta)$ has the same dimension as the operator algebra itself, we are back to the original problem defined by the constraints \eqref{eq:positivity} and \eqref{eq:cons} which gives the correct exact solution $E$ and the corresponding expectation values $\Tr(\rho {\cal O})$ would be the set of physical data associated with the solution (the expectation values that implicitly define $\rho$ as an element of the dual vector space).

A natural question to ask is what is the speed of convergence of $E_*\to E$ for some operator size basis, assuming that the operators that build the basis are natural in some way. This is a much more fine-grained question than the nested inequalities $E_{*,small} \leq E_{*,bigger}\leq E$. What we would like to find is how far we have to go to be inside a bound $E-E_*< \epsilon$, where we choose $\epsilon$ at the beginning. We then want to use that information to understand the complexity of determining $E$ itself. Even more than $E$ as a value, we want to find the approximations of the expectation values $M$ as we increase the size of the basis of operators ${\cal O}_\alpha$. These also eventually converge, but this problem is much harder than how we approximate $E$, as at least in the case of $E$ we have the monotonicity of nested inequalities, whereas no such nested inequality is obvious for the components of the matrix $M$.

Proofs of rapidity of convergence for this type of semidefinite programming problems are hard to produce. In some bootstrap problems related to matrix models and one dimensional statistical measures, one can show exponentially fast convergence on the size of the problem \cite{Berenstein:2025itw,Berenstein:2026wky}, meaning that the errors on the target data can be shown to decrease exponentially with the size of the matrix $M$ that one needs to use to find bounds.

Let us state the problem for a spin chain with $N$ sites. Let us assume that the Hilbert space at each site is $d$-dimensional.
This means that the Hilbert space dimension grows as $d^N$ and the operator algebra grows as $d^{2N}$. We assume that the Hamiltonian is local, so that $H=\sum_i h_i$ where $h_i$ is a local expression on $k$ sites starting at site $i$.
For each $N$, a problem of matrices of size $d^{2N}$ will have converged to the exact answer for $N$ finite.

There are various steps we need to define the actual problem we want to solve. We want to study the thermodynamic limit $N\to \infty$, and the energy density $\langle h_i\rangle= h$, which we assume is translation invariant. Our goal is to minimize $h$
given positivity constraints. This is the particular problem described by \cite{Lawrence:2021msm}.
Translation invariance can always be assumed because of the symmetry arguments of \cite{Scheer:2024eyu}. At worst, we obtain a mixed density matrix that has the same ground state energy. These symmetry constraints always simplify the problem from the point of view of their computational cost. They add constraints that reduce the number of dimensions (free parameters) of the finite SDP problem to solve.

In the thermodynamic limit the quantity $h$ will remain finite, while $E$ grows linearly with $N$. Also, notice that a computer will only have a maximal size of the lattice $N$, but we want to take $N$ to be as large as reasonable. Once we have the energy of the ground state, what we are interested in finding is the correlators of the ground state wave function.

The  problem outlined above  is equivalent to minimizing $E= \sum_i \langle h_i\rangle$ given translation invariance. Our question has two parts. First, how quickly does $h_N$ converge as $N\to\infty$? In a gapped system we expect this to become exponentially fast in $N$, but this assumes that we have determined $h_N$ exactly. For a spin chain at criticality, we only expect polynomial convergence $N^{-a}$ as $N$ becomes large. The second question is about how convergence improves as we increase the size of the basis of ${\cal O}$. This is basis dependent as seen in examples \cite{Lawrence:2021msm,georgeberensteinmeoldpaper}, so there is no simple answer here.
Finally, we would like to estimate the errors on the $M_{\alpha\beta}$ themselves. A way to do so is to relax $E_*$ a little bit (assuming uniqueness), and then any of the problems $\min(M_{\alpha\beta}),\max(M_{\alpha\beta}) $ at fixed $E^*$ is also a convex optimization problem. Such setups give rise to certifiable bounds in quantum systems \cite{Wang:2023hss}, but depending on the quantity that is being bound, the bounds can be rather weak.
We expect each of these to be a soluble semidefinite program problem as well. However, notice  that this procedure of estimating errors is computationally expensive: we need to run an SDP solver for every quantity we want to bound. What we would really want instead is a theoretical understanding of the bounds. Unfortunately, we cannot say much about this problem except in cases where the original SDP is fairly trivial. We will pick up this problem later on in the paper when we study particular examples with free fermion realizations.

\section{Model systems and results}\label{sec:models}

In this work we will focus primarily on two systems. The transverse field Ising model and the 3-state Potts model. We define the Ising model as follows
\begin{equation}
    H=-\sum_{i=1}^N Z_iZ_{i+1}+\mu X_i
\end{equation}
where $Z_i$ and $X_i$ are the usual Pauli matrices (at each site $i$). The 3-state Potts model is defined in terms of the operators $U_a$ and $V_a$ at each site $a$, which satisfy:
\begin{equation}
\begin{aligned}
U_a^3 &= \mathbf{1}, \quad V_a^3 = \mathbf{1} \\
U_a V_a &= \omega V_a U_a \quad \text{where } \omega = e^{2\pi i/3} \\
[U_a, V_b] &= 0 \quad \text{for } a \neq b
\end{aligned}
\end{equation}
In the explicit basis $\{|0\rangle, |1\rangle, |2\rangle\}$:
\begin{align}
U &= \begin{pmatrix} 1 & 0 & 0 \\ 0 & \omega & 0 \\ 0 & 0 & \omega^* \end{pmatrix} = \begin{pmatrix} 1 & 0 & 0 \\ 0 & e^{2\pi i/3} & 0 \\ 0 & 0 & e^{-2\pi i/3} \end{pmatrix}, \quad
V = \begin{pmatrix} 0 & 0 & 1 \\ 1 & 0 & 0 \\ 0 & 1 & 0 \end{pmatrix}
\end{align}

Note that $V$ is a cyclic permutation matrix, while $U$ is diagonal with phases. The Hamiltonian has  a very similar form as the Ising Model
\begin{equation}
    H=-\left(\sum_{i=1}^N U_iU^{-1}_{i+1}+\mu V_i\right)+\text{h.c.} .
\end{equation}
In both cases we can use periodic boundary conditions identifying site $i+N$ with site $i$. The basis used for the SDP for Ising differs slightly from \cite{georgeberensteinmeoldpaper} as in that paper complex Jordan-Wigner fermions were used while in this paper we use all Majorana fermions as defined in Eq.~\ref{eq:majoranadef}. Similarly in the 3-state Potts model case we use the parafermion basis (without numerical coefficients) as defined in Eqs.~\ref{eq:leftparadef}, \ref{eq:rightparadef}, all one point correlators and energy correlators (the Hamiltonian terms). In certain cases we use an extended basis where two-point correlators are added.

Models with more general $q$ were studied by exact diagonalization at $\mu=1$ in \cite{Berenstein:2023ric} and it was noted that they are conformal field theories at $c=1$ with radius $R=\sqrt{2q}$ for $q>4$. This was done by noticing that exactly at that value one has a non-invertible symmetry realized in the field theory (in that case the clock models are motivated by the constructions based on staggered bosons \cite{Berenstein:2023tru}, which seem to automatically produce systems at criticality with non-invertible symmetries realized directly in the local algebra). 
The case $q=3$ is special, because not only is $c<1$, but at criticality it is also integrable \cite{FRADKIN19801} and has a representation in terms of a 
Temperley-Lieb algebra. In that sense, it is a simple model to check if partial information from integrability is effective to solve the spin chain with SDP methods.

\subsection{$\mathbb{Z}_q$ charge constraint}

\begin{table}[h]
\centering
\begin{tabular}{c|cc}
\hline
$N$ & $E_0/N$ & $E_1/N$ \\
\hline
4  & $-1.306563$ & $-1.207107$ \\
6  & $-1.287901$ & $-1.244017$ \\
7  & $-1.283989$ & $-1.251796$ \\
10 & $-1.278491$ & $-1.262751$ \\
\hline
\end{tabular}
\caption{Ground state $E_0$ and first excited state $E_1$ energy densities ($E/N$) for the $\mu=1$ transverse-field Ising model. As seen in \cite{georgeberensteinmeoldpaper} the SDP results agree with the exact spectrum of the ground state to numerical precision. The first excited state is obtained here by imposing $\langle \prod_{i=1}^{N} X_i \rangle = -1$, projecting the SDP into the charged sector. The closing of the gap, $(E_1-E_0)$, with $1/N$ behavior is consistent with a gapless system.}
\label{tab:ising}
\end{table}


The Ising and Potts Hamiltonians have a well known and easily identifiable $\mathbb{Z}_q$ symmetry where $q$ is the number of spin states of the system. This means there are $q$ charge sectors for each theory. In the lattice the corresponding operator is the spin flip operator $\prod_i X_i$ in Ising and $\prod_i V_i$ in Potts (we will use $\prod \sigma$ to refer to both). The eigenvalues correspond to the charge sectors of the theory. As we discovered in \cite{georgeberensteinmeoldpaper} the ground state approached by the system for the vanishing transverse field, $\mu=0$, is a superposition of the $\mathbb{Z}_q$ charges. At criticality we do not expect this behavior as the ground state should fall into a single charge sector. This raises the question whether we can constrain the system such that we can find the ground state of the remaining charge sectors. The simplest constraint one can imagine is setting $\langle\prod\sigma\rangle=\Omega^n$ where $\Omega=e^{2\pi i/q}$ for $n\in [1,...,q) \in \mathbb{Z}$. We see that solely this constraint is satisfactory for calculating the first excited state (ground state of the charge block $n=1$). One way to understand this behavior is that the fermion (parafermion) basis of operators has operators with products similar
to the ${\mathbb Z}_q$ symmetry, so that algebraic relations between the ${\mathbb Z}_q$ operator and other operators can appear together in the positive  matrix $M_{ij}$ we are trying to bound. That way the charge constraint is visible to the rest of the problem.
Otherwise, the $\mathbb{Z}_q$ charge is non-local and local operators do not have enough reach to see a non-local operator. In that sense, we can only fix the charge if the operator basis for the SDP problem is sufficiently non-local.

We correctly get a degeneracy in the $\mathbb{Z}_3$ case as expected (i.e., the constraint associated with $n=1\text{ or }2$ should give the same ground state energy for a charged state). The accuracy is quite high in both cases, however with growing $N$ the differentiation between the two sectors becomes much more difficult to determine as the gap scales with $1/N$. Furthermore for the Potts model, in order to increase the $N$ for which the gap is visible, it is recommended to include constraints such as $\langle\mathcal{O}\prod\sigma\rangle=\langle\mathcal{O}\rangle\cdot\Omega^n$ that specify the symmetry sector. The results are shown in table \ref{tab:potts}. Notice that in the last line we cannot differentiate numerically between the two sectors any longer.
The difference between exact and bootstrap is also zero for very small sizes because the bootstrap problem essentially contains all operators at such small lattices.

\begin{table}[h]
\centering
\begin{tabular}{c|cccc|cccc}
\hline
 & \multicolumn{4}{c|}{Ground state} & \multicolumn{4}{c}{First excited} \\
$N$ & $E_0/N$ & $E_0^{\text{exact}}/N$ & $|\Delta E_0|/N$ & \% error & $E_1/N$ & $E_1^{\text{exact}}/N$ & $|\Delta E_1|/N$ & \% error \\
\hline
3 & $-2.561533$ & $-2.561533$ & $0.000000$ & $0.0000$ & $-2.308000$ & $-2.308000$ & $0.000000$ & $0.0000$ \\
4 & $-2.507425$ & $-2.505475$ & $0.001950$ & $0.0778$ & $-2.365996$ & $-2.365150$ & $0.000846$ & $0.0358$ \\
5 & $-2.488400$ & $-2.480120$ & $0.008280$ & $0.3339$ & $-2.395927$ & $-2.391000$ & $0.004927$ & $0.2061$ \\
6 & $-2.480162$ & $-2.466500$ & $0.013662$ & $0.5539$ & $-2.412654$ & $-2.404933$ & $0.007720$ & $0.3210$ \\
7 & $-2.475143$ & $-2.458357$ & $0.016786$ & $0.6828$ & $-2.475143$ & $-2.413257$ & $0.061886$ & $2.5644$ \\
\hline
\end{tabular}
\caption{Ground state and first excited state energy densities ($E/N$) for the $\mu=1$ 3-state Potts model as a function of lattice size $N$, compared to exact values. The same SDP procedure used for the Ising case is applied here to the (degenerate) charge sectors of the Potts model, with the additional operator constraints $\langle (\prod_i V_i)\, O \rangle = \Omega^n\langle O \rangle$ becoming more effective—and increasingly necessary—at larger $N$. Because the gap closes as $1/N$, even small deviations from the exact excited-state energy can produce complete overlap with the ground state, as observed at $N=7$ where the SDP first excited state energy result coincides with the ground-state value.}
\label{tab:potts}
\end{table}
\subsection{Additional local conserved  charge constraint}

In Minkowski space, the  boost operator is a symmetry that forms part of the Poincare algebra. On the lattice we do not have Lorentz invariance and thus do not have boost charge symmetry. Nonetheless, in many integrable theories, a boost operator can be constructed that does not have a conserved charge itself but serves as a ladder operator between conserved charges \cite{boostbalasz,Bargheerboost,Pozsgay2020boost}. In the case of Ising and the Potts models at criticality there is a tower of conserved charges \cite{infsetconserchargising, Fendley_2014}. The boost operator on the lattice is strictly speaking only defined for infinite chains of integrable systems. However, it can be utilized as an intermediate step to generate the additional conserved charge densities. These can be summed and constrained to generate a conserved charge on a periodic lattice. For a Hamiltonian $H$, the extended charges $Q_n$ are constructed iteratively through commutator relations with a boost operator $B$ and the charge densities are local operators.

Given a continuum definition $B=\int_{-\infty}^\infty x h(x)dx $ one defines an approximate lattice analogue $B= \sum_{j=-\infty}^{\infty} j \cdot H_{j}$ where $H_{j}$ is the self-dual Hamiltonian density at site $j$. For Ising, this is written as $B_{\text{Ising}} = \sum_{j=-\infty}^{\infty} (j+\frac{1}{2})Z_jZ_{j+1}+j\cdot X_j$. Then using the definition from the continuum one defines a tower of conserved charges as follows:
\begin{equation}
\begin{aligned}
Q_2 &= H \\
Q_{j+1} &= -i[B, Q_j] 
\end{aligned}
\end{equation}
\begin{figure}
    \centering
    \includegraphics[scale=.45]{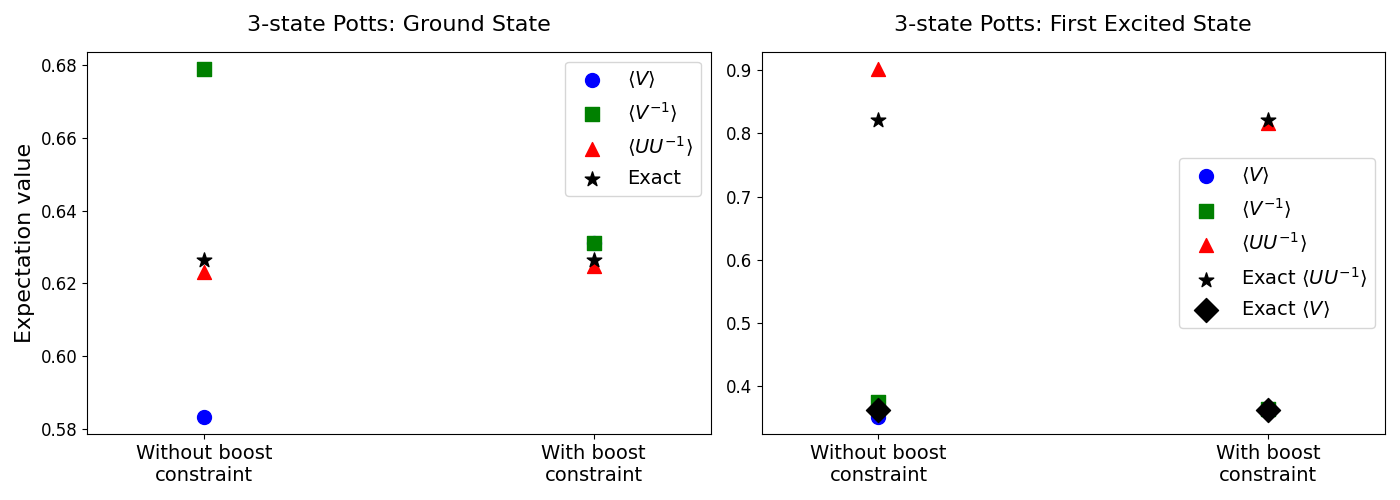}
    \caption{The addition of the additional charge constraints generated from the boost operator have an impact on the SDP results for the Potts model but not the Ising model. In the plot, for simplicity, we call these boost constraints.  Importantly in both models the  charge does not affect the objective function. However we see in these figures that it affects particular correlators. For $N=10$ we see on the left the effect of the  charge constraint on the non invertible symmetry operators in the ground state. The additional charge significantly further constrains them towards the exact answer. On the right we look at the same correlators in the excited state and see similar improvement. No effect was noted in other correlators (e.g., $\langle XX\rangle, \langle ZZ\rangle$, $\langle UU\rangle$ etc.) in either Ising or Potts.}
    \label{fig:potts-boost-charge}
\end{figure}
Note that the boost operator can only be defined in the  infinite lattice, so we need to do the algebra first in the infinite lattice and then restrict to a finite system.
In integrable systems these charges satisfy $[H, Q_n] = 0$ for all $n$, ensuring conservation. 

To incorporate $Q_n$ charge conservation into our semidefinite programming (SDP) approach for ground state optimization, we enforce the constraint that the ground state density matrix $\rho$ satisfies:
$$\langle [Q_n, O] \rangle = \text{Tr}(\rho [Q_n, O]) = 0$$ for all operators $O$ in our basis and all conserved charges $Q_n$. For each conserved charge $Q_n$ and each operator $O_j$ in the SDP basis, we calculate the commutator $[Q_n, O_j] = Q_n O_j - O_j Q_n$ and then  express the result as a linear combination of basis operators: $[Q_n, O_j] = \sum_k c_{jk}^{(n)} O_k$. Numerically, we add the linear constraint: $\sum_k c_{jk}^{(n)} x_k = 0$ where $x_k$ are the SDP variables corresponding to $\langle O_k \rangle$.

Each product calculation $Q_n O_j$ and $O_j Q_n$ is simplified using algebraic relations before forming the commutator. Only constraints where all commutator terms are present in the SDP basis are included. Complex coefficients arising from Pauli algebra (factors of $\omega$, $i$, $-i$, $-1$) also need to be carefully incorporated. This approach ensures that the optimized state respects the conservation laws of the integrable system, providing physically meaningful ground state approximations. From \ref{fig:potts-boost-charge} we see a meaningful effect solely in the Potts model where the non-invertible symmetry is made more accurate in the ground state. The $V$ and $UU$ terms are  more accurate in the excited state as well. We noted no effect in Ising. We will explain why this is the case later on: in the Ising model with the Fermion basis we get the exact ground state anyhow, regardless of the extra constraints. It is natural that there is no improvement in the objective function if we are already at the correct value.


\subsection{Virasoro algebra}

At criticality, a one-dimensional quantum spin chain is expected to flow in the continuum to a $2$D conformal field theory, and the corresponding finite-size spectrum is organized by representations of two commuting copies of the Virasoro algebra~\cite{Virasoro1970}. The universal Casimir correction $-c/(12N)$ to the ground-state energy on a finite circle~\cite{BloteCardyNightingale1986,Affleck1986}, has long been the workhorse for numerical extractions of the central charge. The more ambitious goal of constructing lattice analogues of the Virasoro generators themselves was originally pursued for integrable models via quantum-group techniques~\cite{PasquierSaleur1990}, and put on a more general footing by the proposal that discrete Fourier modes of the local Hamiltonian density furnish, in the scaling limit, the combinations $L_n + \bar{L}_{-n}$ \cite{KOO1994459}. Recent applications of this lattice-Virasoro prescription have produced increasingly precise extractions of CFT data from microscopic models, including the identification of primary fields and OPE coefficients in the critical Ising chain~\cite{ZouMilstedVidal2020}, the detection of emergent Kac-Moody symmetry in critical spin chains with continuous internal symmetry~\cite{WangZouVidal2022}, and the algebraic analysis of logarithmic CFTs arising from supersymmetric lattice models~\cite{GainutdinovReadSaleur2016}. We import this machinery into the SDP setting: rather than diagonalizing the Hamiltonian, we ask whether the SDP optimum, augmented with the appropriate operators, is consistent with the algebraic identities that the lattice Virasoro generators must satisfy at criticality. This provides both a diagnostic for the location of the continuous phase transition and, in principle, a route to extracting CFT data from the SDP correlators.
At criticality, we expect to see a conformal field theory in the continuum. For a finite lattice, we can hope to get some approximation of the conformal algebra results for the spectrum of states. 

The Koo-Saleur formula constructs lattice representations of Virasoro generators through a discrete Fourier transforms: $H_n = \frac{N}{2\pi} \sum_{j=1}^{N} h_j e^{2\pi i n j / N}$ where $h_j$ is the local Hamiltonian density \cite{KOO1994459,guifrekoo}. At criticality, $H_n \propto L_n +\bar{L}_{-n}$ (Virasoro generators). For the critical Ising Model $H_n^{\text{Ising}} = \frac{N}{2\pi} \sum_{j=1}^{N} \left[ e^{2\pi i n (j+1/2)/N}(Z_j Z_{j+1}) + \mu \cdot e^{2\pi i n j/N} X_j \right]$.
\begin{figure}
    \centering
    \includegraphics[scale=.5]{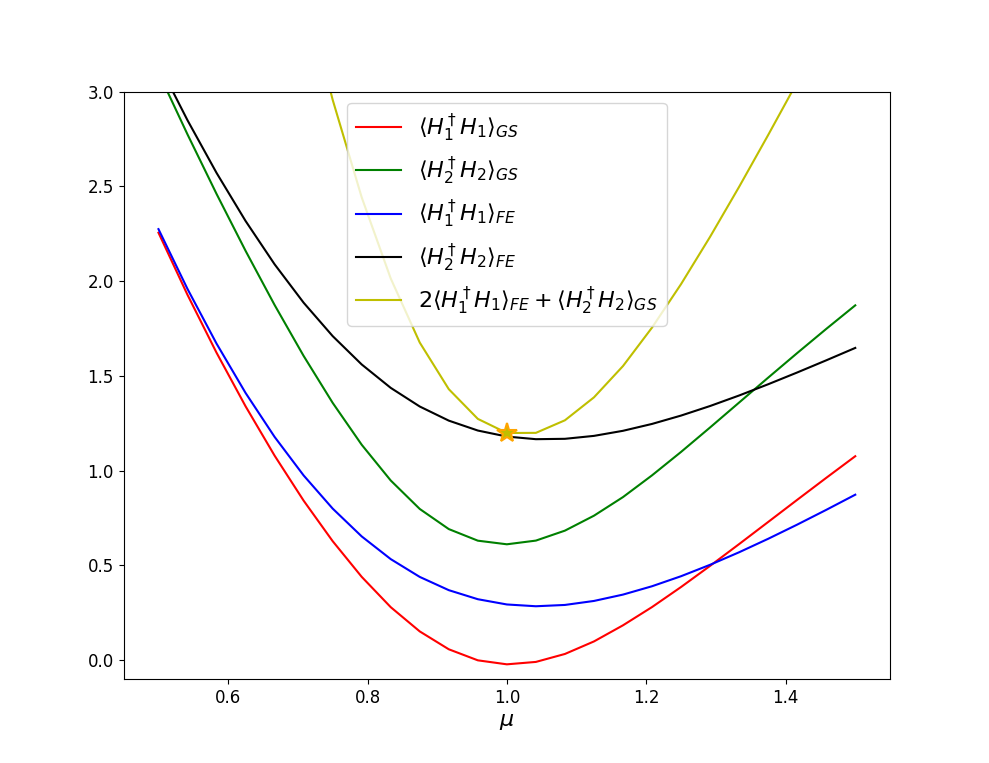}
    \caption{Building up the Virasoro generators for the Ising model allowed us to explore if there was any more CFT data we could determine. Here $FE$ and $GS$ indicate expectation values of the first excited and ground state respectively. For $N=10$ from the $\langle H_1^\dagger H_{1}\rangle_{GS}$ we see that it behaves as expected and approaches 0 at the critical point defined on the horizontal axis as a function of transverse field $\mu=1$. Regardless of normalization or central charge $2\bra{FE}H_1^\dagger H_1\ket{FE}+\bra{I}H_2^\dagger H_2\ket{I}=\bra{FE}H_2^\dagger H_2\ket{FE}$ should hold at criticality which we mark with the yellow star.}
    \label{fig:ising-h1h1}
\end{figure}
At the critical point, the ground state $|0\rangle$ satisfies the conformal vacuum condition $L_n|0\rangle = 0$ for $n > 0$ while $\langle H_n^\dagger H_n \rangle > 0$ away from criticality. A minimum identifies a numerical approximation to the phase transition. We use charge constraints as indicated above to constrain as best we can to $\ket{FE}$ and $\ket{I}$ for the first excited and ground state respectively. Note that one cannot construct the $H_n$ operators solely from the Majorana basis we have been using. One must add the missing cross terms (resulting from multiplying $H_1^\dagger H_1$) to the basis. We also note that adding more terms beyond the missing terms improves the results as well. This includes two point functions of the Pauli operators as well as long strings $\{\prod_{i=1}^{N\geq j>i} Z_i,\quad\prod_{i=1}^{N\geq j>i} Y_i\}$. This suggests that even in the Majorana basis and a free fermion system the SDP is still not learning the whole system as was hinted by the failed correlators in \cite{georgeberensteinmeoldpaper} and which we will continue to note in the coming sections. 

Using the SDP we are able to identify rough features that represent this behavior.  For Ising we see at criticality that $\bra{I} H_1^\dagger H_1\ket{I}=0$ indicating that the SDP has understood that there is no level 1 descendant of the conformal ground state. We can also verify further CFT behavior by checking the relation $2\bra{FE}H_1^\dagger H_1\ket{FE}+\bra{I}H_2^\dagger H_2\ket{I}=\bra{FE}H_2^\dagger H_2\ket{FE}$, which we see is respected at criticality (see Fig.~\ref{fig:ising-h1h1}). This relation is the correct linear combination of the central charge and the conformal weight that arises in the second level descendant of the non-trivial primary.

Using knowledge of the Virasoro algebra approximation we can attempt to go further by using the following knowledge, $\bra{FE} H_1 H_1\ket{FE}=\alpha\Delta_{FE}$ and $\bra{FE} H_2^\dagger H_2\ket{FE}=\alpha(2\Delta_{FE}+\frac{c}{2})$ where $\alpha$ is an unknown scale factor and $\Delta_{FE}$ is the scaling dimension of the first excited state. We also know that $E_1-E_0=\frac{\alpha\Delta_{FE}}{N}$. Now if the SDP approaches the correct ground state the result of this calculation is by necessity $\langle I| H_2^\dagger H_2|I\rangle=\alpha c/2$ where $c$ is the central charge. Given that we already addressed the calculation of the central charge in a previous paper \cite{georgeberensteinmeoldpaper}, here we wish to see if we can use this relation to extract higher energy states.  Putting everything into terms of $c$ thereby removing $\alpha$ gives a scaling dimension of $\Delta_{FE}=.12$ which is qualitatively close to the known continuum answer of .125.

The way we are supposed to interpret this result is that we have an SDP diagnostic for the location of a possible continuous phase transition, plus some additional algebraic checks that the corresponding minima are happening at the same place in the coupling parameter space and that the results can be used to calculate critical exponents (normalized coefficient relations between expectation values in various states identified by symmetry). 

We performed the above procedure in the Potts model and we found the same feature, $\langle H_1^\dagger H_1\rangle\Rightarrow0$, as $\mu\rightarrow 0$ indicative of the phase transition  near the critical point. This information can be found in figure \ref{fig:paravira} in the Appendix. It required a much larger basis of operators, but the procedure seems effective there as well. Other CFT quantities like the central charge, did not work particularly well.

Now, if we want to perhaps force the SDP to discover the descendant $L_{-2}\ket{I}=\sqrt{c/2}\ket{T}$ where $\ket{T}$ is the spin-2 quasi-primary stress tensor state then we can perhaps use the relation

\begin{equation}\label{eq:h2t}
\begin{aligned}
\bra{T}H_2^\dagger H_2\ket{T}&=2\langle I| H_2^\dagger H_2|I\rangle\\
\bra{\bar{T}}H_2^\dagger H_2\ket{\bar{T}}&=(\frac{16}{c}+2)\langle I| H_2^\dagger H_2|I\rangle
\end{aligned} 
\end{equation}
where $L_0\ket{I}=0\ket{I}$ (see the rest of the calculation in Appendix). Unfortunately we were unable to constrain the energy of the optimal of the SDP to be within $15\%$ of the correct $\bra{T}H\ket{T}$. While we hoped to be able to use the Virasoro to consecutively generate and constrain spin states we concluded that this is beyond the accuracy that the SDP can provide. 


One should also note that one could attempt to extract the value of descendant correlators through the following relationship which is solely reliant on solving the SDP for the ground state.
\begin{equation}
\bra{\psi_n} O \ket{\psi_n}
\;=\; \frac{\bra{I}\,H_{-n}^\dagger\, O\, H_{-n}\,\ket{I}}
            {\bra{I}\,H_{-n}^\dagger H_{-n}\,\ket{I}}
\end{equation}
However this necessitates adding new states to the basis so that all operators are present for the calculation. Clearly one can only generate descendants of a primary state (either the ground state or one accessed using other constraints such as charge mentioned earlier). It is important to note again that the SDP still only finds the ground state of the system $|I\rangle$ and not the $\psi_n$. If $O=H$ bounds on energy of excited states within the Verma module can be set. One can also imagine using this as a way to constrain the system either for all $O$ or instead just implementing the constraint $\bra{\psi_n} H \ket{\psi_n}=E_{\psi_n}$ and seeing the response of the correlators.
\section{Fermionic basis and free fermions}\label{sec:free}

For a Gaussian (free fermion) state, Wick's theorem states that any correlation function of fermion operators can be expressed as a sum over all possible pairwise contractions. For four fermion operators, the complete Wick's theorem expression is:

\begin{equation}\label{eq:wt}
\langle c_i c_j c^\dagger_k c^\dagger_l \rangle = \langle c_i c^\dagger_k \rangle \langle c_j c^\dagger_l \rangle - \langle c_i c^\dagger_l \rangle \langle c_j c^\dagger_k \rangle + \langle c_i c_j \rangle \langle c^\dagger_k c^\dagger_l \rangle
\end{equation}

The minus sign between the first two terms comes from fermionic signs when we reorder operators. In order to check the behavior of our state we use fermionic variables. In the Ising model we can use the Jordan-Wigner fermion operators, which are well known 
\begin{equation}\label{eq:jwdef}
\begin{aligned}
c_i &= \frac{1}{2} \prod_{j=0}^{i-1} [X_j] (Y_i - iZ_i) \\
c_i^\dagger &= \frac{1}{2} \prod_{j=0}^{i-1} [X_j] (Y_i + iZ_i).
\end{aligned}
\end{equation}

\begin{figure}
    \centering
    \includegraphics[scale=.35]{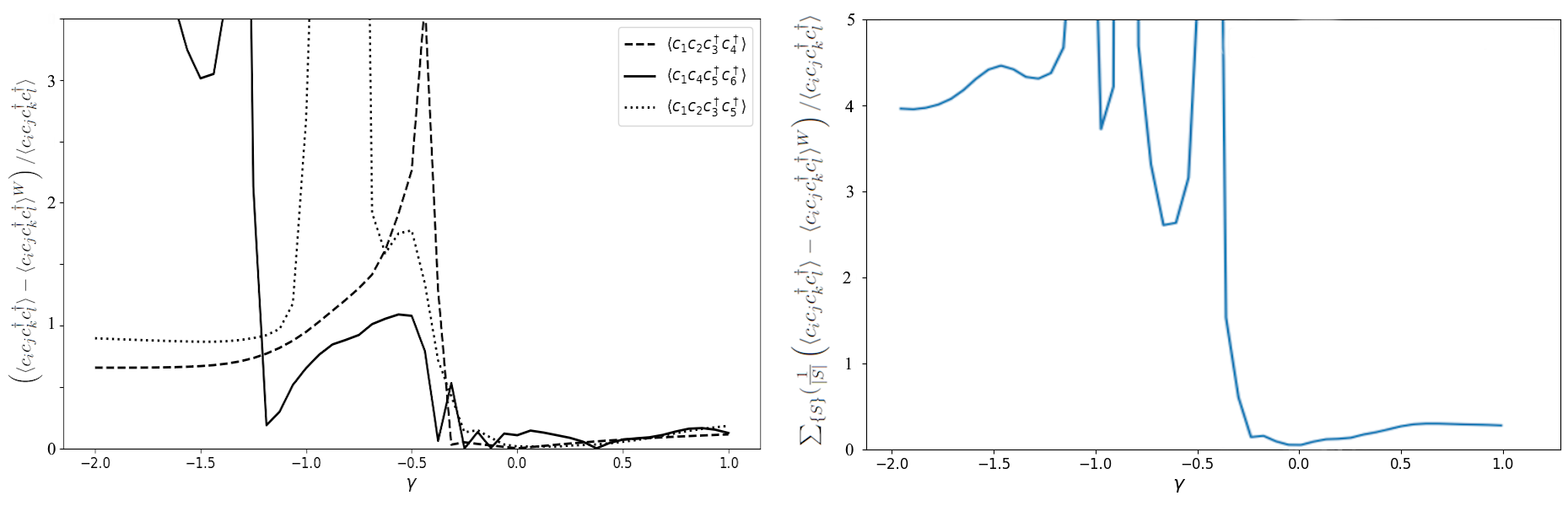}
    \caption{The left figure is the ratio, $\left(\langle c_i c_j c_k^\dagger c_l^\dagger \rangle-\langle c_i c_j c_k^\dagger c_l^\dagger \rangle^W\right)/\langle c_i c_j c_k^\dagger c_l^\dagger \rangle$, where $\langle c_i c_j c_k^\dagger c_l^\dagger \rangle$ is calculated directly from multiplying all four fermion operators and $\langle c_i c_j c_k^\dagger c_l^\dagger \rangle^W$ is the right hand side of Wick's theorem (Eq.~\ref{eq:wt}). The horizontal axis is $\gamma$ from Eq.~\ref{eq:annni} for $N=10$. The right figure is, $\sum_{\{S\}}(\frac{1}{|S|}\left(\langle c_i c_j c_k^\dagger c_l^\dagger \rangle-\langle c_i c_j c_k^\dagger c_l^\dagger \rangle^W\right)/\langle c_i c_j c_k^\dagger c_l^\dagger \rangle$, where the sum is over the elements $\{i,j,k,l\}\in S$ where $S$ is the set of all fermion four point functions equivalent to multiplications of fermion two point functions at the level of the Pauli matrices. $|S|$ is the dimension of the set thus making the graph a representation of the average adherence or violation of Wick's theorem. As was the case when looking at fourier modes of the Hamiltonian, one can access more states with an enlargement of the basis set to improve performance. This is done by adding two point functions and long strings $\{\prod_{i=1}^{N\geq j>i} Z_i,\quad\prod_{i=1}^{N\geq j>i} Y_i\}$. While the SDP does seem to approach a qualitative understanding of Wick's state it does not reproduce it in every scenario accurately even with the fermion basis. Qualitative reasoning shows that many of the left hand side four point functions are not present in the basis because the moment matrix is a result of the multiplication of two point functions. Putting in the known basis elements of the four point functions would be necessary. We note a feature that in the literature it is known that Eq.~\ref{eq:annni} has an emergent Ising CFT for $-0.3 < \gamma < 250$ \cite{guifrekoo}. This seems to coincide with the strong violations of Wick's theorem in the plots towards negative values $\gamma<-0.3$.}
    \label{fig:wtandwtavg}
\end{figure}

By constructing the operators based on this transformation we can test Wick's theorem as expressed in Eq.~\ref{eq:wt}. Furthermore we can try to identify what is the behavior of the emergent CFTs in the ANNNI model \cite{KOO1994459}:
\begin{equation}\label{eq:annni}
H_{\text{ANNNI}} = \sum_{i=1}^{N} Z_i Z_{i+1} + \gamma  Z_i Z_{i+2} + X_i + \gamma  X_i X_{i+1}
\end{equation}
In Fig.~\ref{fig:wtandwtavg} we see that at $\gamma=0$ we recover  Wick's theorem for some correlation functions. It is necessary to look at various four point functions as the percent errors at the Ising point can vary from less than one percent to 15\% or more. Importantly not all four point functions are available for testing because of the nature of the basis we used: it generates all two point functions we need, but not all four point functions. The SDP does not know about the four point function decomposition especially at greater distances. We would have to add the four point functions into the solver which means that the scaling with system size $N$ would suffer substantially from making the SDP problem variables scale like a higher power of $N$. In that sense, we avoid it to keep the size (algorithmic complexity) of the problem under control.
We should also note that we can use Wick's theorem not only to study if the state is a free fermion. Strong violations of Wick's theorem in the graph suggest a connection with phase transitions in the model. 
At strong coupling, although the system preserves the non-invertible symmetry, it is expected that the system will be a gapped topological phase with a degeneracy which is a multiple of three \cite{Seiberg:2024gek}.

\subsection{Majorana Fermions}
A further natural description of the Ising system comes in the form of Majorana Fermion variables defined as $\beta_i=c_i+c_i^\dagger$ and $\gamma_i=-i(c_i-c_i^\dagger)$. In terms of these operators the Hamiltonian has a particularly simple form 
\begin{equation}\label{eq:majBH}
    H=\sqrt{-1}\left(\sum_{i=1}^{N}\beta_i\gamma_{i+1}+\beta_i\gamma_i\right)
\end{equation}
Importantly for later discussion note that the Majorana Fermion variables anticommute with each other and square to one $\{\beta_i,\beta_j\}=2\delta_{ij}$,$\{\gamma_i,\gamma_j\}=2\delta_{ij}$, $\{\beta_i,\gamma_j\}=0$,  and that 
\begin{equation}\label{eq:majoranadef}
\begin{aligned}
\beta_i &= Y_i \prod_{k=1}^{i-1} X_k \\
\gamma_i &= Z_i \prod_{k=1}^{i-1} X_k
\end{aligned}
\end{equation}
\begin{figure}
    \centering
    \includegraphics[scale=.5]{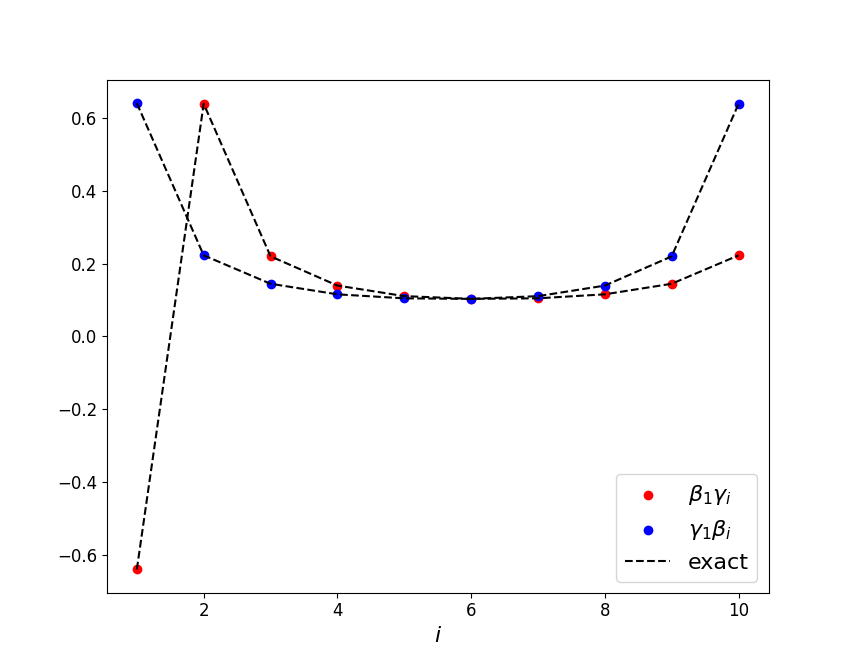}
    \caption{Here we plot $\langle \beta_1\gamma_i\rangle$ and $\langle\gamma_1\beta_i\rangle$ for a $N=10$ lattice. The circles represent the SDP results while the dashed line is from exact diagonalization. Note the correct anticommutation behavior for $\{\beta_1,\gamma_1\}=0$. From this graph the suggestive shift, $\langle\gamma_1\beta_j\rangle=\langle\beta_1\gamma_{j+1}\rangle$, of the correlators suggests the redefinition $\beta_j=\gamma_j$ and $\gamma_j=\beta_{j-1}$. This redefinition is a manifestation in the Majorana variables of the Kramers-Wannier duality of the model.}
    \label{fig:maj2p}
\end{figure}

In Fig.~\ref{fig:maj2p} we see the behavior of selected two point functions of the Majorana variables. From the simple analysis we are able to verify the correct anticommutation behavior and the Kramers-Wannier duality of the model.
As we see, the semidefinite program  seems to get the exact answer for the correlators. Our next goal is to explain why this is actually a correct statement.

\subsection{How the SDP solves the free fermion systems.}
Consider the following general linear combination of Majorana fermion operators
\begin{equation}
\psi=\sum c_j \theta_j
\end{equation}
where the $c_j$ are complex. The operator $\psi\psi^\dagger\succeq 0$ is positive. 
Consider the expectation values $M_{i,j}=\langle \theta_i \theta_j \rangle$. Using antisymmetry and the hermiticity of $M$  (derived from positivity of $\psi^\dagger\psi$), we find that $M_{i,i}=1$, $M_{i,j}=-M_{j,i} $ and that the off-diagonal $M_{i,j}$ is purely imaginary. We can write this as follows
\begin{equation}
M= {\bf 1} + i A
\end{equation}
where $A$ is an antisymmetric matrix.
Since $A$ is antisymmetric, its eigenvalues come in pairs $i \lambda, -i \lambda$. Diagonalizing $M$, or equivalently, $A$, we find that positivity of $M$ is equivalent to $1\pm \lambda \geq 0$. If we focus on the positive $\lambda\geq0$, then $\lambda\leq 1$.

Let us assume that we have a quadratic Hamiltonian $\hat H= \sum \alpha_{ij} \theta_i \theta_j$. The expectation value can then be written as an expression $\langle\hat H \rangle= \sum {\alpha_{ij}} M_{ij} \simeq \Tr( \alpha \cdot M)$.
Again, $\min(\langle\hat H \rangle)|_{M\succeq 0}$ is a standard SDP problem. Since the variable to be optimized is linear, consider two interior points where $M$ is strictly positive. It is easy to show that $a M_1+(1-a) (M_2)$ is positive when $1\geq a\geq 0$.
This positivity can extend a little further if neither $M_1, M_2$ have zero eigenvalues. One can easily show that given this information, the objective function $\langle\hat H \rangle$ is linear in $a$ and that the minimum will occur at the edge of positivity (the maximum or minimum value $a$ where the matrix $a M_1+(1-a) (M_2)\succeq 0$ is still positive semidefinite). What this means, is that at the minimum $M$ has at least one zero eigenvalue, or equivalently, $A$ has at least one eigenvalue equal to $i$. Each such eigenvalue implies that there is a pair of raising/lowering operators, such that the lowering operator annihilates the vacuum. That is, one can factorize a free fermion. The argument can be repeated in the smaller Hilbert space where said fermion is removed, so one can find that there is a second free fermion and so on. That is, the 
SDP will find that $M$ is in a corner of positivity where there is a maximum possible number of zero eigenvalues: the SDP must therefore find all the free fermion raising/lowering operators and therefore, implicitly, the ground state that is annihilated by all the lowering operators.

Let us explain this argument in detail. Notice  that when we have a decomposition into pairs of eigenvalues $i\lambda, -i\lambda$, the split into positive and negative eigenvalues can also be associated with a pair of raising/lowering operators. One can say that the split defines a complex structure as follows: consider an eigenvector of $A$, $\vec v$ and decompose it into real and imaginary part $\vec v = \vec v_1+i \vec v_2 $. Since $A\vec v = i\lambda \vec v$, we find $A\vec v_1=\lambda v_2$ and $A\vec v_2=-\lambda v_1$. It is easy to check that $A (\vec v_1-i \vec v_2)=-i \lambda (\vec v_1-i \vec v_2)$, so that the pair of eigenvectors related by complex conjugation $\vec v, \vec v^*$ have opposite eigenvalues with respect to $A$. For generic $A$ (so long as it is even dimensional), there are no zero eigenvalues. In this way, the basis $\vec v_{\lambda>0}$ provides anti-holomorphic coordinates of the vector space, and the complex conjugates provide holomorphic coordinates. This is equivalent to a split into raising and lowering operators, where the raising operator is holomorphic and the lowering operator is anti-holomorphic.

In that setup, we can further show that for that specific $A$, there is at least one density matrix $\rho$ such that $M_{ij}=\langle \theta_i \theta_j\rangle_\rho$. That is, there exists at least one density matrix that satisfies all the constraints compatible with konwing $M$.

The idea follows from diagonalizing $A$. For a value of $\lambda$ that is not maximal, on the corresponding free fermion associated with $\lambda$,  we can  realize the expectation value by the mixed state $ a\ket 1 \bra 1 +(1-a) \ket 0 \bra 0 $ for the corresponding pair of raising/lowering operators, where $\lambda \equiv 1-2a$. 
This is in keeping with fermions having zero expectation value $\langle \theta_i\rangle=0$.
What this means is that to each element of the set $M\succeq 0$ one can associate a factorization into fermion pairs (raising/lowering operators) and a corresponding factorized density matrix (a type of generalized Gibbs density matrix), 
where
\begin{equation}
    \rho= \bigotimes_{\lambda >0} \rho_\lambda 
\end{equation}
Each of these is also a Gaussian (Wick) state: Wick's theorem is satisfied
for each such $\rho$. 

What is important is that if the Hamiltonian is a free fermion model, for each $M\succeq 0$, there exists at least one density matrix $\rho$ as described above. The minimization problem can also be thought of as a variational problem where we minimize over $\rho$ as defined above. This way, the solution of the SDP will produce an upper bound on the energy, that is,  $E_*\geq E_0$.  
On the other hand, we argued that $M\succeq 0$ must be satisfied by unitarity, so that the standard version of the SDP method we have advocated must produce a number $E\leq E_0$. Putting these together, we must get the exact value of the ground state energy (we can call this a sandwich argument). 
Moreover, when we evaluate $M$ at the extremum,  it will also implicitly find the factorization. When $\lambda=1$, the state for each such factorization is pure and therefore the corresponding density matrix is that of a pure state. That is, barring degeneracies, the SDP method produces a solution that has a unique density matrix of a pure state $\rho$ associated with the data that it finds.

Consider the more general problem when we deform the system away from free fermions. Any time $M\succeq 0$ has an eigenvalue equal to $0$ ( $A$ has a pair $\pm i$), we must factor out a free fermion (this is strictly when $M$ is  given only by the bifermions by $M=\langle \theta_i\theta_j\rangle$). Notice that if we have a more general interacting field theory, we don't expect that a free fermion can be factorized in the ground state of the system. What this implies is that all the eigenvalues of $M$ are larger than zero: the matrix $M$ is at an interior point of the convex set rather than at the edge. In that sense, the inequalities arising from  $M\succeq 0$ become ineffective: the problem will have a lot of freedom to move away from the correct value while keeping positivity. 

What this means is that when the program produces a solution, $M_*$ can be very different than $M_{true}$.
This can only be fixed if we add more information, for example, by using more complicated fermionic operators $\theta_i, \theta_i \theta_{i+1}\theta_{i+2}$ in the basis (see the basis choices in \cite{Scheer:2024eyu}). Effectively, we must increase the information in the matrix $M$, so that the new matrix $\tilde M$ constraints the problem better. In this case, we cannot prove convergence to the correct value $E_0$, nor that a putative $\rho_M$ exists for each $M$ that satisfies the enlarged constraints. Some of the statements we have made above will generalize to this setup. For example, the fact that the general $M$ (not just $M$ restricted to bifermions) that solves the problem will lie on the boundary of the positivity region. This implies (by being careful to extract the null eigenspace of $M$) that there exists at least one operator ${\cal O}$ such that $\langle {\cal O}^\dagger {\cal O} \rangle=0$, or equivalently ${\cal O}\ket 0= 0$ annihilates the state. In a fully interacting field theory such operators would be akin to lowering operators for dressed particles and these are not expected to be simple operators. 
The operator ${\cal O}$ that we find this way is an approximation to such dressed operators and we expect corrections when we include more information. Only in the free case are such operators  expected to be simple.

Now let us see how our results apply in the example of the Ising problem. 
In this case the SDP optimization starts with Pauli correlator variables
$$\mathcal{V} = \{\langle X_i \rangle, \langle Y_i \rangle, \langle Z_i \rangle, \langle \sigma^a_i \sigma^b_j \rangle, \langle X_0 X_1 Z_2 \rangle,\ldots,\langle X_0 X_1 Y_2\rangle\ldots\},$$
where the objective function to minimize is
$$E = -J \sum_{i=1}^{N} \langle Z_i Z_{i+1} \rangle - \mu\langle X_i \rangle$$
and the constraint is that $M\succeq 0$ where 
$M_{ij}=\mathcal{V}_{i}\cdot\mathcal{V}_j$ where the indices iterate over all entries in $\mathcal{V}$. $M_{ij}$ are thus the variables in the problem. Let us focus on the Majorana fermion variables block section of $M$. For $N=4$ and a Majorana two point correlator $u_{ij}=\chi_i^\dagger\chi_j$ the block we focus on is as follows 
\begin{equation}\label{matrix:M}
M
=
\begin{pmatrix}
u_{00} & u_{01} & u_{02} & u_{03} \\
u_{10} & u_{11} & u_{12} & u_{13} \\
u_{20} & u_{21} & u_{22} & u_{23} \\
u_{30} & u_{31} & u_{32} & u_{33}
\end{pmatrix}
=
\begin{pmatrix}
1 & u_{01} & u_{02} & u_{03} \\
- u_{01} & 1 & u_{12} & u_{13} \\
- u_{02} & - u_{12} & 1 & u_{23} \\
- u_{03} & - u_{13} & - u_{23} & 1
\end{pmatrix}
=
\mathbf{1} + i A.
\end{equation}
We are also imposing translation invariance as if there was no end to the chain, so that $u_{01}=u_{12}=u_{23}$. If the spin chain has periodic boundary conditions for the fermions, we would also have $u_{30}=u_{01}$, whereas if it is antiperiodic, we would have $u_{30}=-u_{01}$. Both of these give periodic boundary conditions for the energy density.

What we are
actually minimizing is the local density at one point in the lattice $\langle  -JZ_0 Z_1-\mu X_1\rangle $, so we are not imposing the periodicity of the boundary conditions ab initio. 
We are also imposing other discrete symmetries that make it possible to set some correlators to zero to lower the computational cost of the problem (this can always be done \cite{Scheer:2024eyu}).
Given that we don't input the boundary conditions,  what does the program actually do?  It will seek for a solution where the correlator $M$ is in a corner and that also has periodic boundary conditions for the energy density correlators. There are two possibilities: periodic or antiperiodic boundary conditions of the fermions. It is known that the second one has less energy, so the system picks that one. That one also has less energy than the thermodynamic limit. One can also argue that the thermodynamic limit will only produce a non-trivial density matrix on a subset, because one is tracing over the region that is not captured by the correlators. In that sense, any such SDP will look  for the boundary conditions that minimize the energy density given all other constraints and that the result must end up in a pure state.

When there are zero modes we saw in \cite{georgeberensteinmeoldpaper} that our SDP approached a superposition state of the possible ground states rather than choosing a single ground state. This usually arises because we work with symmetry constraints and the two possible vacua can break the symmetry. The density matrix that is diagonal in the two (multiple) vacua respects the symmetry and is the one that is found by the symmetry constrained SDP. 

Furthermore we note that while the Majorana fermion description describe excitations across the spin chain, the SDP does not actually use all this information. Note that $i\gamma_j\beta_j=X_j$ and $i\beta_j\gamma_{j+1}=Z_jZ_{j+1}$. Now let's consider calculating $\langle\beta_i \gamma_i \beta_j \gamma_j\rangle$. Remembering Wick's theorem
\begin{equation}
    \langle\beta_i \gamma_i \beta_j \gamma_j\rangle = \langle \beta_i \gamma_i \rangle \langle \beta_j \gamma_j \rangle - \langle \beta_i \beta_j \rangle \langle \gamma_i \gamma_j \rangle - \langle \beta_i \gamma_j \rangle \langle \gamma_i \beta_j \rangle
\end{equation}
we are interested in situations where $i<j$. We know that in the continuum $\langle \beta_i\beta_j\rangle=\delta_{ij}$ and $\langle \gamma_i\gamma_j\rangle=\delta_{ij}$, however they are not necessarily zero in the lattice nor are they set to zero by $\mathbb{Z}_2$ symmetry so let us keep them. Now let's rewrite the above in terms of Pauli matrices
\begin{equation}
\langle X_i X_j \rangle = \langle X_i \rangle \langle X_j \rangle + \left\langle Z_i Y_j \prod_{k=i+1}^{j-1} X_k  \right\rangle \left\langle Y_i Z_j\prod_{k=i+1}^{j-1} X_k  \right\rangle + \left\langle Y_i Y_j\prod_{k=i+1}^{j-1} X_k  \right\rangle \left\langle Z_i Z_j\prod_{k=i+1}^{j-1} X_k  \right\rangle
\end{equation}
where we see that the two point correlator of $X$ at long distance is reflective of correlators of order $i-j$. Now let's consider the two point correlator $\langle Z_iZ_j\rangle$, how can we express this in terms of Majorana fermions? To begin, remember the properties of the Majorana fermions $\{\gamma_m, \gamma_n\}= 2\delta_{mn}, \{\beta_m, \beta_n\} = 2\delta_{mn}, \{\gamma_m, \beta_n\} = 0$ and $X_k = i\gamma_k\beta_k $. Now from $\gamma_i = Z_i \prod_{k=1}^{i-1} X_k$ and using $X_k^2 = I$ we can invert to find
\begin{equation}
Z_i = \gamma_i \prod_{k=1}^{i-1} X_k
\end{equation}
Now for $i<j$
\begin{equation}
Z_i Z_j = \left(\gamma_i \prod_{k=1}^{i-1} X_k\right) \left(\gamma_j \prod_{k=1}^{j-1} X_k\right).
\end{equation}
Remembering that $i<j$ and thus $k<j$ therefore $[\gamma_j, X_k] = [\gamma_j, \gamma_k\beta_k]  = 0$ for $j \neq k$. Since $\gamma_j$ commutes with all $X_k$ for $k < j$,
\begin{equation}
Z_i Z_j = \gamma_i \gamma_j \left(\prod_{k=1}^{i-1} X_k\right) \left(\prod_{k=1}^{j-1} X_k\right)
\end{equation}
which is further simplified given that the strings overlap from $k=1$ to $k=i-1$:
\begin{align}
\left(\prod_{k=1}^{i-1} X_k\right) \left(\prod_{k=1}^{j-1} X_k\right) &= \left(\prod_{k=1}^{i-1} X_k\right) \left(\prod_{k=1}^{i-1} X_k\right) \left(\prod_{k=i}^{j-1} X_k\right) \\
&= \prod_{k=1}^{i-1} X_k^2 \cdot \prod_{k=i}^{j-1} X_k \\
&= \prod_{k=i}^{j-1} X_k
\end{align}
giving us the statement that 

\begin{equation}
Z_i Z_j = \gamma_i \gamma_j \prod_{k=i}^{j-1} X_k= \sqrt{-1}\beta_i\gamma_j\prod_{k=i+1}^{j-1}X_k=(-1)^{\frac{j-i}{2}}\gamma_i \gamma_j \prod_{k=i}^{j-1} \gamma_k\beta_k=(-1)^{\frac{j-i}{2}}\beta_i\gamma_j\prod_{k=i+1}^{j-1}\gamma_k\beta_k
\end{equation}

The failure of the SDP to accurately represent the $\langle X_i X_j\rangle$ correlator is indicative that the model does not have a strong understanding of Wick's theorem. Moreover, $\langle X_i \rangle \langle X_j \rangle$ is usually non-zero, so the connected piece of $\langle X_i X_j\rangle$ is also subject to numerical floating point error from subtraction.
Indeed, at the critical point, the operator  $X-\langle X\rangle$ becomes the energy operator in Ising.

\section{Parafermions}

Parafermions were introduced by Fradkin and Kadanoff~\cite{FRADKIN19801} as a generalization of Jordan-Wigner fermions appropriate to $\mathbb{Z}_q$-symmetric statistical systems, building on earlier disorder-variable constructions~\cite{KadanoffCeva1971}. These do not obey fermionic anticommutation relationships and there is no Wick's theorem decomposition into two point functions. However, the continuum counterparts---the $\mathbb{Z}_q$ parafermion conformal field theories---form one of the canonical families of rational CFTs and have been studied extensively in the context of edge modes of fractional quantum Hall states and topological phases~\cite{AliceaFendley2016}. Unlike Majorana fermions, parafermions satisfy non-trivial $\mathbb{Z}_q$ braiding relations rather than anticommutation, and their utility for solving lattice problems is correspondingly more limited: even when the Hamiltonian is quadratic in parafermions, the model is generically not free. Recent work has clarified both the boundaries and the limits of solvability in this setting, including the integrable structure of the chiral Potts model and its parafermionic descendants~\cite{Baxter2014}, exact results on the spectrum of free-parafermion chains with multispin interactions~\cite{AlcarazPimenta2020}, the conformal invariance of these models in the critical regime~\cite{AlcarazRamos2024}, and the realization of parafermion modes in proposed solid-state platforms~\cite{KlinovajaLoss2014}. We use the parafermion description here as a diagnostic: rather than expecting it to solve (which \cite{georgeberensteinmeoldpaper} suggested it does not) the SDP as the Majorana basis solves Ising, we ask whether the SDP at least respects the parafermion algebra in its optimal solution, which would indicate some structural understanding of the model even in the absence of a free representation.

The left parafermion operators have strings extending to $+\infty$:
\begin{equation}\label{eq:leftparadef}
\begin{aligned}
\hat{\beta}_{L,2a-1} &= \omega V_a^\dagger U_a^\dagger \prod_{j=a+1}^{\infty} U_j^\dagger \\
\hat{\beta}_{L,2a} &= V_a^\dagger \prod_{j=a+1}^{\infty} U_j^\dagger 
\end{aligned}
\end{equation}
The right parafermion operators have strings extending to $-\infty$:
\begin{equation}\label{eq:rightparadef}
\begin{aligned}
\hat{\beta}_{R,2a-1} &=  V_a^\dagger \prod_{j=-\infty}^{a-1} U_j^\dagger\\
\hat{\beta}_{R,2a} &= \omega^2 U_a^\dagger V_a^\dagger\prod_{j=-\infty}^{a-1} U_j^\dagger 
\end{aligned}
\end{equation}
The parafermion operators satisfy:
\begin{equation}\label{eq:paraalgabra}
\begin{aligned}
\hat{\beta}_{L,b}^3 &= \hat{\beta}_{R,b}^3 = 1, \\
\hat{\beta}_{L,b} \hat{\beta}_{L,c} &= \omega^{\text{sgn}(c-b)} \hat{\beta}_{L,c} \hat{\beta}_{L,b}, \\
\hat{\beta}_{R,b} \hat{\beta}_{R,c} &= \omega^{\text{sgn}(b-c)} \hat{\beta}_{R,c} \hat{\beta}_{R,b}, \\
\hat{\beta}_{L,b} \hat{\beta}_{R,c} &= \begin{cases}
\omega^{(-1)^b} \hat{\beta}_{R,c} \hat{\beta}_{L,b} & b = c, \\
\hat{\beta}_{R,c} \hat{\beta}_{L,b} & b \neq c.
\end{cases}
\end{aligned}
\end{equation}

While in the Ising model the decomposition of the problem into fermions was the key to solving the system, that is not the case in the Potts model. Even though the Hamiltonian is quadratic in the parafermions one immediately sees that these parafermions cannot be broken down into raising and lowering operators and do not have a linear relationship, that is,  they don't anti-commute. Furthermore when one considers the parafermions fusion rules, these involve other fields meaning that they cannot factor into two point functions:  there is no close analogy to Wick's theorem. Also,  while in the Ising model the energy field indeed corresponds to simple fermion correlators, here in the parafermion case the energy field does not and in fact does not even have a $\mathbb{Z}_{3}$ symmetry. While at the level of the lattice we impose periodicity we don't impose any such case on the level of the parafermions and so all long chains of operators terminate at site 1 or $N$. The use of this Potts parafermion analysis was done to verify that the parafermion algebra was respected by the SDP results and that the SDP was developing an understanding of the system.

\section{Conclusion}\label{sec:conc}

In this paper we have studied various improvements to SDP methods for solving spin chain Hamiltonian problems and we  have especially focused on the points at criticality. Additional information provided to the SDP  allowed for the determination of energies of charged excited states and was able to qualitatively find the location of some phase transitions using ideas from approximate Virasoro constraints on the lattice.
We also explained why the SDP can solve certain aspects of free fermion models exactly, including the determination of boundary conditions that minimize the energy and the two point correlation functions of the fermions, but higher point functions are challenging as the amount of variables that the SDP needs to get the values of these correlators grows like higher powers of the lattice volume $L^d$.
In problems that are not equivalent to free fermions, the SDP methods struggle to get the correlators precisely, but qualitatively they agree with some exact diagonalization results for small systems.
We see this in the study of the Potts model.
Also, by using the approximate Virasoro in the Potts model we were able to find the location of the critical point to a reasonable degree. Even though that model is integrable at criticality, that was not enough to get the correlators correctly.
 
What is the way forward then? One way to look at the problem is that the SDP methods on their own are very generic and applicable in any quantum system. In that sense, they are always true, but not necessarily the most effective.

By themselves, the SDP methods we have described do not know intrinsically about the physical insights that have been used historically to solve problems in QFT. Ideas  like the renormalization group, the role of entanglement in the ground state properties, what locality means for correlators and
the notion of scales and effective field theory are all absent in the naive formulation of the SDP problems. Some of these are used only a posteriori to analyze the output. 
We can hope that including these ideas in a way that adds additional constraints to the SDP methods might provide better results and be used to effectively solve problems with these methods.

\acknowledgments

We are very grateful to G. Hulsey for discussions. D.B. would like to thank the Institute of Physics at the University
of Amsterdam for their hospitality while this work was being carried out. The work of D.B. was
supported in part by the Department of Energy under grant DE-SC 0011702.

\bibliographystyle{apsrev4-1}
\bibliography{refs}
\appendix

%
%

\appendix

\section{Construction of the boost operator and local conserved charges}\label{app:boos}

In this appendix we give a careful derivation of the lattice boost operator used in the main text, and we show how the coefficients of its local terms are fixed by demanding that the resulting commutator with $H$ produces a legitimate translation-invariant conserved charge density. This second requirement only works on integrable systems however. The motivation for performing this derivation in some detail is twofold. First, the boost operator is, strictly speaking, not defined on a periodic chain: it carries an explicit coordinate $j$ which is incompatible with translation invariance. Our strategy is therefore to work on the infinite lattice, where the boost operator is unambiguous, then to read off the local conserved charge density that it generates and only afterwards impose periodic boundary conditions on that density. Second, the SDP problem only ever uses the local pieces of $Q_n$, so it is the local form of the result, not the boost operator itself, that ultimately enters the optimization. 
We begin by writing the most general translation-covariant ansatz for the boost operator on the infinite Ising chain, parametrized by two unknown coefficient functions $a(i)$ and $b(i)$ for the two types of local term in the Hamiltonian density:
\begin{align}
B&=\sum_{i=-\infty}^{\infty}a(i)Z_iZ_{i+1}+b(i)X_i,\\
H&=\sum_{i=-\infty}^{\infty}Z_iZ_{i+1}+X_i.
\end{align}
The next conserved charge in the integrable tower is defined by
\begin{equation}
Q_3=-i[B,H],
\end{equation}
and computing this commutator amounts to working out the four cross-terms that arise from pairing each piece of $B$ with each piece of $H$:
\begin{align}
Q_3&=-i\sum_{i,j}(a(i)Z_iZ_{i+1}+b(i)X_i)(Z_jZ_{j+1}+X_j)\nonumber\\
&\quad-(Z_jZ_{j+1}+X_j)(a(i)Z_iZ_{i+1}+b(i)X_i)\\
&=-i\sum_{i,j}a(i)Z_iZ_{i+1}X_j+b(i)X_iZ_jZ_{j+1}-b(i)Z_jZ_{j+1}X_i-a(i)X_jZ_iZ_{i+1}.
\end{align}
A small notational comment is in order: in the first line above, the prefactor $-i$ is the imaginary unit, while $i,j$ are lattice indices. To avoid this collision in the subsequent algebra we will denote the imaginary unit by $j=\sqrt{-1}$ from here on, freeing $i$ to play its usual role as a site label.

The sum over $j$ in the expression above produces nonzero contributions only when $j$ coincides with one of the sites in the support of the corresponding $B$ term, since otherwise the Pauli operators commute. Carrying out the sums and grouping by site gives
\begin{align}
Q_3&=-i\sum_ia(i)Z_iZ_{i+1}(X_i+X_{i+1})+(b(i)X_i+b(i+1)X_{i+1})Z_{i}Z_{i+1}\\ \nonumber
&\quad-Z_iZ_{i+1}(b(i)X_i+b(i+1)X_{i+1})-a(i)(X_i+X_{i+1})Z_iZ_{i+1}.
\end{align}
We can now evaluate the remaining commutators between $Z$ and $X$ on the same site, which simply convert each pair into a $Y$ Pauli matrix with an explicit factor of $j$:
\begin{align}
Q_3&=-i\sum_ia(i)(jY_iZ_{i+1}+jZ_iY_{i+1})+(-jb(i)Y_iZ_{i+1}-jb(i+1)Z_iY_{i+1})\\ \nonumber 
&\quad-(b(i)jY_iZ_{i+1}+b(i+1)jZ_iY_{i+1})-a(i)(-jY_iZ_{i+1}-jZ_iY_{i+1}).
\end{align}
The factor of $j$ pulls out of the sum and combines with the overall $-i$ to give a real overall coefficient, leaving
\begin{equation}
Q_3=2\sum_i (a(i)-b(i))(Y_iZ_{i+1})+(a(i)-b(i+1))(Z_iY_{i+1}).
\end{equation}
At this point we impose the physical requirement that the resulting charge density be translation invariant, which is precisely the condition that the coefficients of $Y_iZ_{i+1}$ and $Z_iY_{i+1}$ are equal up to the sign expected from the antisymmetric structure of $Q_3$. Concretely we require
\begin{equation}
a(i)-b(i)=b(i+1)-a(i),
\end{equation}
which together with the natural normalization $b(i)=i$ gives
\begin{equation}
b(i)=i\ \Rightarrow\ a(i)=i+\tfrac{1}{2}.
\end{equation}
That is, the boost operator is built from the lattice Hamiltonian density weighted by $j$ for the $X$ piece and by $j+1/2$ for the $Z_jZ_{j+1}$ piece, recovering the standard offset that reflects the fact that the bond operator naturally lives midway between the two sites it couples. Substituting back yields
\begin{equation}
Q_3=\sum_iY_iZ_{i+1}-Z_iY_{i+1}.
\end{equation}
The same procedure, applied recursively, generates the next charge in the tower:
\begin{equation}
Q_4=-i[B,Q_3]=\sum_i Z_iX_{i+1}Z_{i+2}-Y_iY_{i+1}-Z_iZ_{i+1}-X_i.
\end{equation}
This matches the form of the local conserved charge densities obtained by other methods in the integrable Ising literature~\cite{infsetconserchargising}. To use these inside the SDP on a finite periodic chain we simply discard $B$ and retain only the local densities, summing them with the periodic identification $i+N\equiv i$. The boundary terms that this procedure misses are subleading in $1/N$ and do not affect the SDP constraint $\langle [Q_n,\mathcal{O}]\rangle=0$ for local operators $\mathcal{O}$ in the bulk.

\section{Level-two descendant calculation in the lattice Virasoro algebra}\label{app:yellowmatch}

The purpose of this appendix is to derive the algebraic identity that links the expectation value $\langle T|H_2^\dagger H_2|T\rangle$ to quantities that are already determined by the SDP solution, where $|T\rangle$ denotes the spin-2 quasi-primary stress tensor descendant of the conformal vacuum. The identity is used in Section~\ref{sec:models} as an attempted route to push the SDP into a higher excited state by adding it as an algebraic constraint. 

The starting point is the Koo-Saleur identification $H_n\propto L_n+\bar L_{-n}$, valid at criticality up to an overall normalization $\alpha$ which we will carry as an unknown. The descendant state is defined by $|T\rangle=\sqrt{2/c}\,L_{-2}|I\rangle$, where $|I\rangle$ is the identity (conformal vacuum), normalized so that $\langle T|T\rangle=1$. We also use the standard Virasoro algebra relations $[L_n,L_m]=(n-m)L_{n+m}+\frac{c}{12}n(n^2-1)\delta_{n+m,0}$ together with the fact that the vacuum is annihilated by $L_n|I\rangle=0$ for $n\geq -1$, and the assumption that holomorphic and antiholomorphic generators commute and annihilate the same vacuum.

With these ingredients the calculation proceeds by expanding $H_2^\dagger H_2$ in terms of the chiral generators, repeatedly moving annihilation operators to the right until they act on $|I\rangle$, and using the commutators to pick up Casimir terms whenever an $L_n$ has to be moved past an $L_{-m}$. The step-by-step expansion is as follows.

Using the Koo--Saleur identification $H_2 = \alpha(L_2 + \bar L_{-2})$, so that $H_2^\dagger = \alpha(L_{-2} + \bar L_2)$:

\begin{align}
H_2^\dagger H_2 &= \alpha^2\,(L_{-2} + \bar L_2)(L_2 + \bar L_{-2})\\
&= \alpha^2\left(L_{-2}L_2 + L_{-2}\bar L_{-2} + \bar L_2 L_2 + \bar L_2 \bar L_{-2}\right).
\end{align}

\begin{align}
\bra{I}H_2^\dagger H_2\ket{I}
&= \alpha^2\bra{I}\left(L_{-2}L_2 + L_{-2}\bar L_{-2} + \bar L_2 L_2 + \bar L_2 \bar L_{-2}\right)\ket{I}\\
&= \alpha^2\bra{I}\bar L_2 \bar L_{-2}\ket{I}\\
&= \alpha^2\bra{I}[\bar L_2,\bar L_{-2}]\ket{I}\\
&= \alpha^2\bra{I}\left(4\bar L_0 + \tfrac{c}{2}\right)\ket{I}\\
&= \alpha^2\,\tfrac{c}{2},
\end{align}

since every term with $L_2$ or $\bar L_2$ on the right annihilates $\ket{I}$, and $\bar L_0\ket{I}=0$. Similarly,

\begin{align}
H_2\ket{T}
&= \alpha\sqrt{\tfrac{2}{c}}\,(L_2 + \bar L_{-2})L_{-2}\ket{I}\\
&= \alpha\sqrt{\tfrac{2}{c}}\,\Big(L_2 L_{-2}\ket{I} + \bar L_{-2}L_{-2}\ket{I}\Big)\\
&= \alpha\sqrt{\tfrac{2}{c}}\,\Big(\underbrace{[L_2,L_{-2}]\ket{I}}_{(4L_0+\frac{c}{2})\ket{I}=\frac{c}{2}\ket{I}} + \bar L_{-2}L_{-2}\ket{I}\Big)\\
&= \alpha\sqrt{\tfrac{2}{c}}\,\Big(\tfrac{c}{2}\ket{I} + \bar L_{-2}L_{-2}\ket{I}\Big).
\end{align}

The two resulting states are orthogonal (different antiholomorphic weight), so

\begin{align}
\bra{T}H_2^\dagger H_2\ket{T}
&= \alpha^2\,\tfrac{2}{c}\,\Big(\tfrac{c}{2}\Big)^2\braket{I}{I}
   + \alpha^2\,\tfrac{2}{c}\,\big\|\bar L_{-2}L_{-2}\ket{I}\big\|^2\\
&= \alpha^2\,\tfrac{2}{c}\,\Big(\tfrac{c}{2}\Big)^2
   + \alpha^2\,\tfrac{2}{c}\,\underbrace{\bra{I}L_2\bar L_2\,\bar L_{-2}L_{-2}\ket{I}}_{(c/2)(c/2)}\\
&= \alpha^2\,\tfrac{2}{c}\,\Big(\tfrac{c}{2}\Big)^2 + \alpha^2\,\tfrac{2}{c}\,\Big(\tfrac{c}{2}\Big)^2\\
&= \alpha^2\,\tfrac{2}{c}\cdot 2\Big(\tfrac{c}{2}\Big)^2\\
&= \alpha^2\,c.
\end{align}

Therefore
\begin{equation}
\frac{\bra{T}H_2^\dagger H_2\ket{T}}{\bra{I}H_2^\dagger H_2\ket{I}}
= \frac{\alpha^2 c}{\alpha^2 c/2} = 2
\end{equation}
independent of $c$. Also,

\begin{align}
H_2\ket{\bar T}
&= \alpha\sqrt{\tfrac{2}{c}}\,(L_2 + \bar L_{-2})\bar L_{-2}\ket{I}\\
&= \alpha\sqrt{\tfrac{2}{c}}\,\Big(\underbrace{L_2\bar L_{-2}\ket{I}}_{=\,\bar L_{-2}L_2\ket{I}=0} + \bar L_{-2}\bar L_{-2}\ket{I}\Big)\\
&= \alpha\sqrt{\tfrac{2}{c}}\,\bar L_{-2}^2\ket{I}.
\end{align}

The norm of the level-4 state is obtained from the algebra:
\begin{align}
\bar L_2\,\bar L_{-2}^2\ket{I}
&= \big[\bar L_2,\bar L_{-2}\big]\bar L_{-2}\ket{I} + \bar L_{-2}\bar L_2 \bar L_{-2}\ket{I}\\
&= \big(4\bar L_0 + \tfrac{c}{2}\big)\bar L_{-2}\ket{I} + \bar L_{-2}\big(4\bar L_0 + \tfrac{c}{2}\big)\ket{I}\\
&= \big(8 + \tfrac{c}{2}\big)\bar L_{-2}\ket{I} + \tfrac{c}{2}\,\bar L_{-2}\ket{I}\\
&= (8 + c)\,\bar L_{-2}\ket{I},
\end{align}
where in the first term $\bar L_0$ acts on the weight-2 state $\bar L_{-2}\ket{I}$ (giving $4\cdot 2 = 8$), while in the second it acts on the vacuum (giving $0$). Hence
\begin{equation}
\big\|\bar L_{-2}^2\ket{I}\big\|^2 = \bra{I}\bar L_2^2\,\bar L_{-2}^2\ket{I}
= (8+c)\,\bra{I}\bar L_2\bar L_{-2}\ket{I} = (8+c)\,\tfrac{c}{2}.
\end{equation}

Therefore
\begin{align}
\bra{\bar T}H_2^\dagger H_2\ket{\bar T}
&= \alpha^2\,\tfrac{2}{c}\,\big\|\bar L_{-2}^2\ket{I}\big\|^2\\
&= \alpha^2\,\tfrac{2}{c}\,(8+c)\,\tfrac{c}{2}\\
&= \alpha^2\,(8 + c),
\end{align}
and
\begin{equation}
\frac{\bra{\bar T}H_2^\dagger H_2\ket{\bar T}}{\bra{I}H_2^\dagger H_2\ket{I}}
= \frac{\alpha^2(8+c)}{\alpha^2 c/2} = \frac{2(8+c)}{c} = \frac{16}{c} + 2
\end{equation}
which depends on $c$.

\section{Koo-Saleur diagnostics in the Potts model}\label{app:koosaleurpotts}

In the main text we showed that, at criticality, the Ising SDP supports a clean interpretation of the lowest Fourier modes of the Hamiltonian density as approximate lattice Virasoro generators, and that the associated expectation values exhibit the expected behavior of a level-one null state in the ground sector together with a recognizable ratio between $H_2^\dagger H_2$ in the ground and first-excited states. It is natural to ask whether the same diagnostic survives when transplanted to the three-state Potts model, where the continuum CFT is the $c=4/5$ minimal model and the lattice model is integrable only exactly at $\mu=1$. This appendix reports the result of that numerical experiment.

The methodology is identical to the Ising case. We construct the Fourier-transformed Hamiltonian density $H_n$ via the Koo-Saleur prescription, use charge constraints to isolate the ground and excited sectors of the SDP, and evaluate $\langle H_n^\dagger H_n\rangle$ as a function of the transverse field $\mu$. The only differences from the Ising setup are the parafermion basis we work in and the absence of a free representation of the model, which means that the operator algebra closing under the SDP basis is more restrictive. We probed two basis choices: a minimal basis consisting of the parafermion operators in Eqs.~\ref{eq:leftparadef}--\ref{eq:rightparadef}, the Hamiltonian terms, and all one-point functions; and an enlarged basis that additionally includes all two-point functions at distances $N$ and $N/2$.

\begin{figure}
    \centering
    \includegraphics[scale=.5]{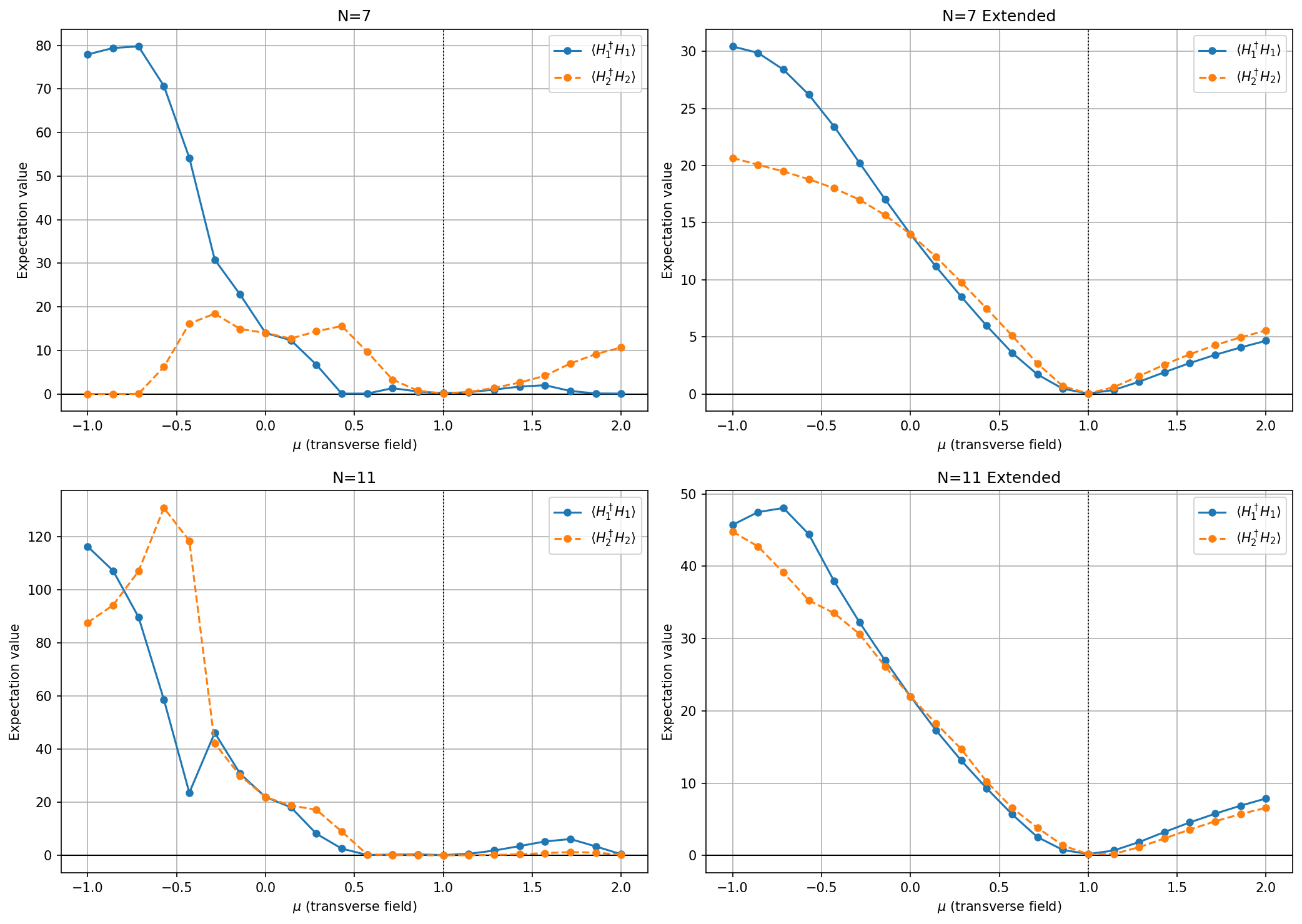}
    \caption{$\langle H_n^\dagger H_n\rangle$ for the three-state Potts model as a function of the transverse field $\mu$, using the parafermion basis of Eq.~\ref{eq:rightparadef} together with the Hamiltonian terms and all one-point functions (left), and an enlarged basis that additionally includes all two-point functions at distances $N$ and $N/2$ (right). The enlarged basis produces a recognizable feature near the critical coupling, similar to  Fig.~\ref{fig:ising-h1h1}. Notice also that the same quantity can receive very large corrections from one basis to the next.}
    \label{fig:paravira}
\end{figure}

Figure~\ref{fig:paravira} shows the resulting $\langle H_n^\dagger H_n\rangle$ curves for both choices. The minimal basis fails to produce a clear minimum at the known critical coupling. The enlarged basis is  better, exhibiting a localized feature in the vicinity of the critical point, but the feature is shallower than the corresponding Ising signal and shifted by a small but visible amount from the exact location of the transition. Neither basis reproduces the level-one null condition $\langle I|H_1^\dagger H_1|I\rangle=0$ to anywhere near the precision we achieved in Ising, and the ratio identities involving $H_2^\dagger H_2$ that worked at criticality in Ising fail to hold in any quantitative sense.

Our reading of these results is that the failure is structural rather than numerical. The Koo-Saleur identification of $H_n$ with lattice Virasoro generators depends on the local Hamiltonian density behaving, in the continuum limit, like the energy-momentum tensor of the underlying CFT. In Ising this works because the SDP already has access to a free representation of the system in the Majorana basis, so the operators that appear in $H_n^\dagger H_n$ are well constrained even after the SDP relaxation. In Potts the parafermion basis does not factor the system, so the four-point and higher correlators implicit in $H_n^\dagger H_n$ are not fixed by the two-point block, and the SDP has enough latitude in those correlators to smooth out the algebraic structure that the Koo-Saleur prescription is supposed to expose. Enlarging the basis with more long-range two-point functions partially recovers the diagnostic, but the cost scales unfavorably with $N$, and starts to become an ``art form" of what operators to add. One can say this is consistent with the general theme of the paper that, away from free fermions, extracting structural information from SDP correlators is exponentially expensive in lattice size.

\section{Parafermion two-point functions in the SDP solution}\label{app:parafermionsdp}
In this section the basis used is the parafermion basis defined in Eq.~\ref{eq:leftparadef},\ref{eq:rightparadef} as well as their complex conjugates and all one point functions and energy correlators. The two families of lattice parafermions, $\hat{\beta}_{2a-1}$ and 
$\hat{\beta}_{2a}$ (and their right, left definitions), exhibit qualitatively distinct two-point correlation 
functions, motivating their separate analysis. For the odd family 
$\hat{\beta}_{2a-1}$, 
the same-chirality correlators 
$\langle \hat{\beta}^\dagger_{1,R}\, \hat{\beta}_{2a-1,R}(a) \rangle$ 
and 
$\langle \hat{\beta}^\dagger_{1,L}\, \hat{\beta}_{2a-1,L}(a) \rangle$ 
are predominantly real and decay rapidly from unity at $a=1$, 
consistent with a correlator dominated by a single local operator at 
short distance. The SDP bootstrap values agree closely with exact 
diagonalization for these same-chirality cases, demonstrating that the 
operator basis adequately constrains them. 

\begin{figure}
    \centering
    \includegraphics[scale=.4]{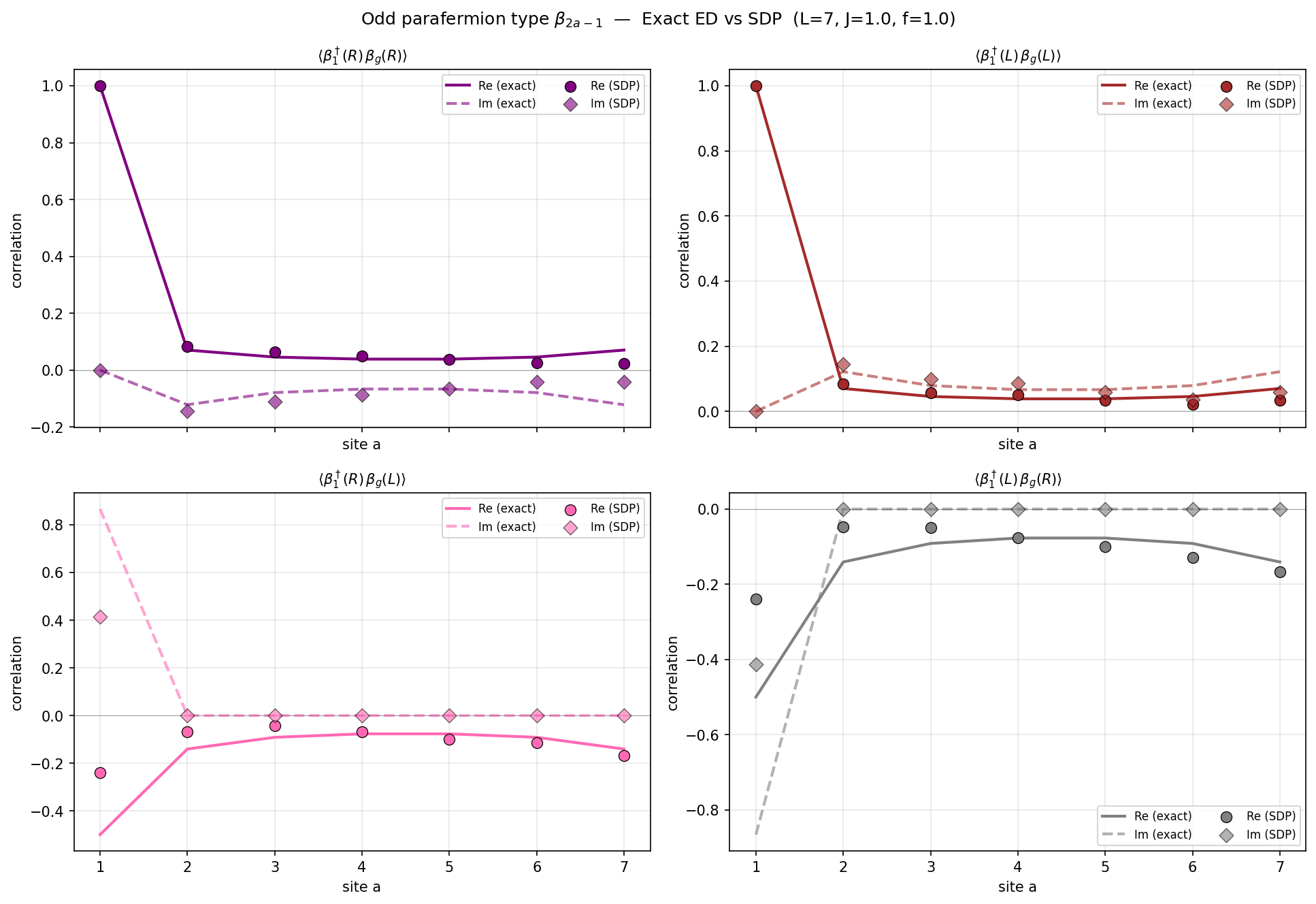}
    \caption{Two-point correlators 
    $\langle \hat{\beta}^\dagger_{1,\chi}\, \hat{\beta}_{2a-1,\chi'}(a) \rangle$ 
    for the odd parafermion family $\hat{\beta}_{2a-1}$
    as a function of site $a$, for a periodic three-state Potts chain of length $L=7$ 
    at the self-dual critical point. Solid and dashed lines show exact 
    diagonalization results for real and imaginary parts respectively; circles and 
    diamonds show the corresponding SDP bootstrap values. The same-chirality 
    correlators (top row) show strong agreement between SDP and exact diagonalization, 
    with the real part decaying rapidly from unity at $a=1$ and the imaginary part 
    remaining near zero throughout the chain. The mixed-chirality correlator 
    $\langle \hat{\beta}^\dagger_{1}(L)\, \hat{\beta}_{2a-1}(R) \rangle$ (bottom right) 
    shows the largest discrepancy between SDP and exact results: the SDP imaginary 
    part collapses to zero while the exact imaginary part is large, indicating this 
    correlator is poorly constrained by the current operator basis.}
    \label{fig:odd_parafermion}
\end{figure}

\begin{figure}
    \centering
    \includegraphics[scale=.4]{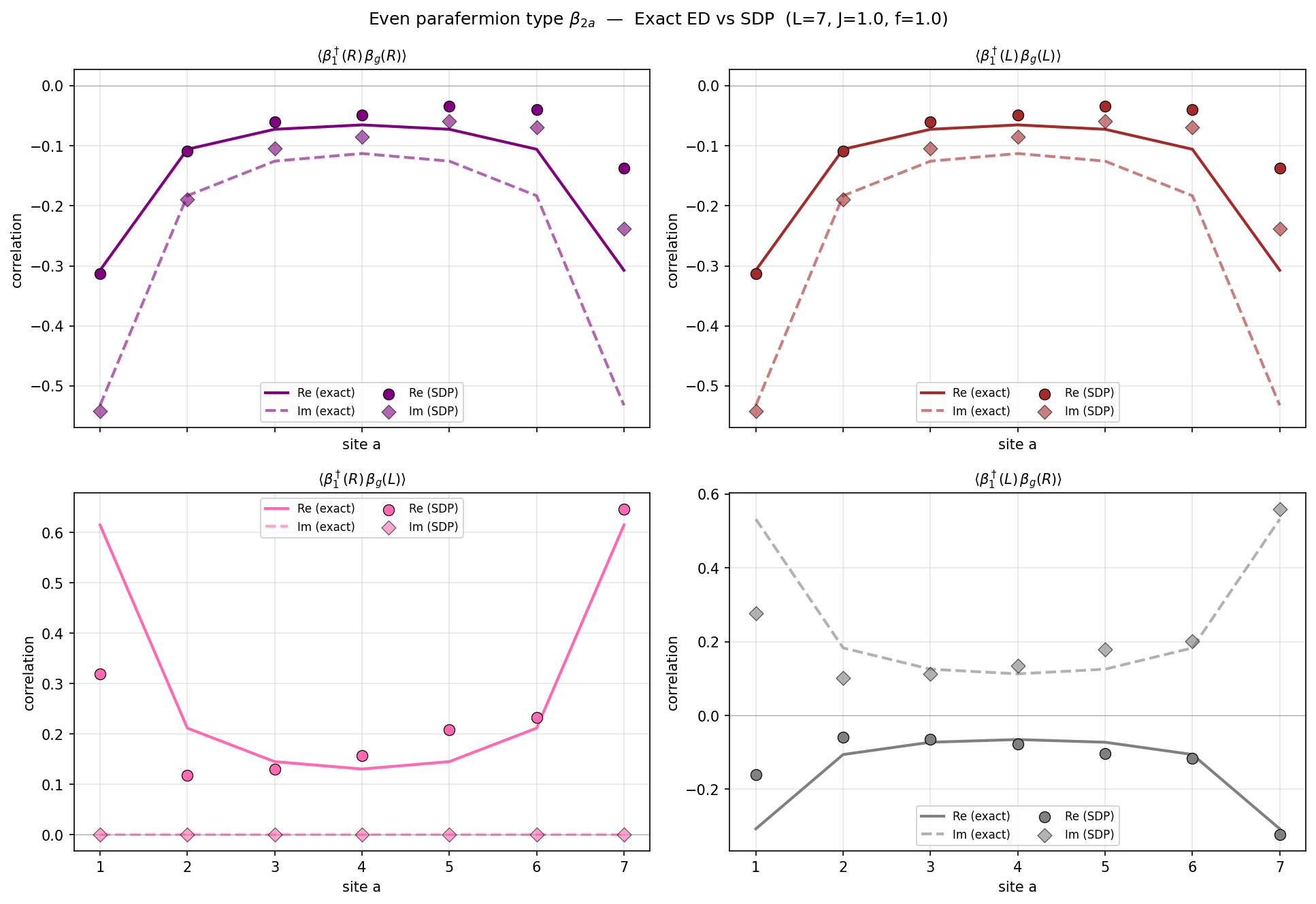}
    \caption{Two-point correlators 
    $\langle \hat{\beta}^\dagger_{1,\chi}\, \hat{\beta}_{2a,\chi'}(a) \rangle$ 
    for the even parafermion family $\hat{\beta}_{2a}$ as a function of site $a$, 
    for the same system as Figure~\ref{fig:odd_parafermion}. In contrast to the 
    odd family, the real parts are negative and exhibit a smooth dome-shaped 
    profile, peaking near the centre of the chain and falling toward the 
    boundaries --- a characteristic finite-size effect of an open-boundary 
    parafermion string on a periodic chain. The imaginary parts are large and 
    structured, reflecting the nontrivial phase carried by the $\omega^2$ 
    prefactor in the $\hat{\beta}_{2a}$ definition. The correlator 
    $\langle \hat{\beta}^\dagger_{1}(R)\, \hat{\beta}_{2a}(L) \rangle$ 
    (bottom left) has a U-shaped real part growing toward the boundaries, 
    while its imaginary part is essentially zero in both exact and SDP results. 
    SDP agreement is good for the same-chirality correlators and reasonable for 
    the mixed-chirality ones, with systematic deviations at the boundary 
    sites $a=1$ and $a=7$.}
    \label{fig:even_parafermion}
\end{figure}

For the even family $\hat{\beta}_{2a}$, the 
correlators have a markedly different character. The real parts are 
negative and dome-shaped, reflecting the interplay between the 
open-boundary parafermion string and the periodic Hamiltonian: the 
string operators for $\hat{\beta}_{2a}$ extend further across the chain 
than those of $\hat{\beta}_{2a-1}$, making boundary effects more 
pronounced. The large imaginary parts arise from the $\omega^2$ prefactor 
in the definition of $\hat{\beta}_{2a}$, which rotates the correlator 
into the complex plane. Together, Figures~\ref{fig:odd_parafermion} 
and~\ref{fig:even_parafermion} demonstrate that the SDP bootstrap 
correctly captures the same-chirality parafermion correlations across 
both families, while mixed-chirality correlators, particularly those 
involving $\hat{\beta}^\dagger(L)$ paired with $\hat{\beta}(R)$, 
necessitate an extended operator basis in order to improve.

\end{document}